\documentclass[12pt]{article}
\input epsf
\usepackage{graphicx}
\usepackage{bm}
\usepackage{latexsym}

\newcommand{\bea}{\begin{eqnarray}}
\newcommand{\eea}{\end{eqnarray}}
\newcommand{\be}{\begin{equation}}
\newcommand{\ee}{\end{equation}}
\newcommand{\pa}{\partial}
\newcommand{\nn}{\nonumber \\}
\newcommand{\e}{\epsilon}
\newcommand{\ve}{\varepsilon}
\newcommand{\w}{\omega}
\newcommand{\Tr}{\mbox{Tr}}
\newcommand{\tr}{\mbox{tr}}
\newcommand{\hH}{\hat{H}}
\newcommand{\hPhi}{\hat{\Phi}}
\newcommand{\hPi}{\hat{\Pi}}
\newcommand{\tPhi}{\tilde{\Phi}}
\newcommand{\tD}{\tilde{D}} 

\makeatletter

\@addtoreset{equation}{section}
\makeatother

\def\href#1#2{#2}

\begin{document}

\begin{titlepage}
\vspace*{30mm}

\begin{center}
{\LARGE \sl Confined Phase In The Real Time Formalism And \vspace{8mm}\\ 
The Fate Of The World Behind The Horizon}\\
\vspace*{15mm}
{\sl FURUUCHI \ Kazuyuki}\\
\vspace*{3mm}
{\sl Harish-Chandra Research Institute}\\
{\sl Chhatnag Road, Jhusi, Allahabad 211 019, India}\\
{\tt furuuchi@mri.ernet.in}
\vspace*{4mm}
\begin{abstract}
In the real time formulation 
of finite temperature field theories,
one introduces an additional set of fields (type-2 fields)
associated to each
field in the original theory 
(type-1 field).
In \cite{Maldacena:2001kr},
in the context of the AdS-CFT correspondence,
Maldacena interpreted type-2 fields as
living on a boundary 
behind the black hole horizon.
However,
below the Hawking-Page transition temperature,
the thermodynamically preferred configuration is
the thermal 
AdS without a black hole, and hence
there are no horizon and boundary behind it.
This means that when the dual gauge theory is
in confined phase, the type-2 fields cannot be
associated with the degrees of freedom behind the
black hole horizon.
I argue that in this case 
the role of the type-2 fields 
is to make up
{\it bulk} type-2 fields of
classical closed string field theory
on AdS at finite temperature in
the real time formalism.
\end{abstract}
\end{center}

\setcounter{footnote}{0}
\end{titlepage}       

\tableofcontents
\section{Introduction}

The AdS-CFT correspondence \cite{Maldacena:1997re}
has provided us with powerful tools to
tackle the puzzles surrounding black holes 
in asymptotically 
Anti-de Sitter (AdS) spaces.
Strong evidences for correspondence between
deconfinement phase transition
in gauge theory and Hawking-Page transition
to black hole geometry \cite{Hawking:1982dh}
have been given in 
\cite{Witten:1998qj,Witten:1998zw}
in the Euclidean path integral
formulation of finite temperature field theory
\cite{Matsubara:1955ws,Gibbons:1976ue}
(also referred to as imaginary time formalism).
Again in the imaginary time formalism,
the author recently showed
how the expectation value of the Polyakov loop,
the order parameter of confinement-deconfinement
transition,
encodes the dual bulk geometry  
in correlation functions of
gauge invariant operators 
\cite{Furuuchi:2005qm}.\footnote{%
See also 
\cite{Brigante:2005bq} for further 
investigation
on some aspects of the confined phase in
large $N$ gauge theories in
the imaginary time formalism, and 
\cite{Furuuchi:2005eu} for
its relation to the large $N$ reductions.}
However, the real problems
about black holes,
such as
the information loss paradox,
dynamical formation and evaporation
of black holes, causal structure and 
singularities,
can only be studied
in the Lorentzian signature.

\begin{figure}
\begin{center}
 \leavevmode
 \epsfxsize=100mm
 \epsfbox{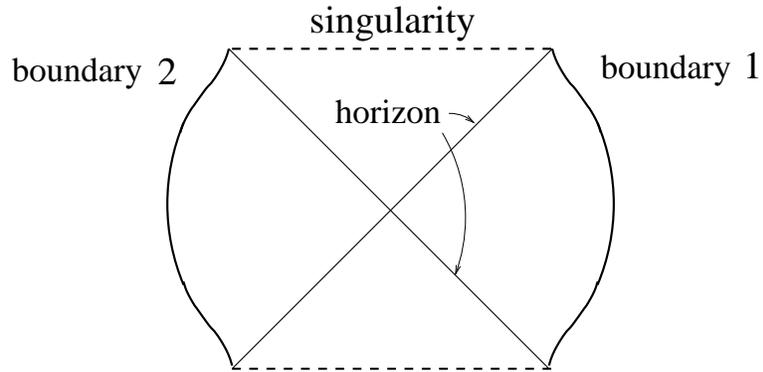}\\
\end{center}
\caption{The extended Carter-Penrose diagram
of the AdS-Schwarzshild black hole geometry.
Spherical directions are suppressed in the figure.
Besides the usual boundary of the AdS space
at spatial infinity (boundary 1, see also Fig.\ref{AdS}),
there is a second boundary behind the horizon (boundary 2).
For a more detail, see e.g. \cite{Fidkowski:2003nf}.}
\label{AdSBH}
\end{figure}

In the real time formulation of
finite temperature field theories
\cite{%
Takahasi:1974zn,Umezawa:1982nv,Semenoff:1982ev,%
Niemi:1983nf,Niemi:1983ea},\footnote{%
See \cite{Niemi:1983nf} 
and a review \cite{Landsman:1986uw}
for the real time formulation of
finite temperature field theories
relevant for this article.}
it seems unavoidable
to introduce 
an additional set of fields besides
the original ones
(those in the zero-temperature theory). 
Each of the newly introduced field
is associated with a field in the original theory.
(In this article I will call newly introduced
fields as type-2 fields, as opposed 
to the original fields
which I will call type-1 fields.)
On the other hand, the
extended Carter-Penrose diagram
of the AdS-Schwarzshild black hole geometry
has a boundary behind the horizon
(Fig.\ref{AdSBH} boundary 2),
in addition to the usual 
boundary of the AdS space at spatial
infinity outside the horizon 
(Fig.\ref{AdSBH} boundary 1).
\begin{figure}
\begin{center}
 \leavevmode
 \epsfxsize=30mm
 \epsfbox{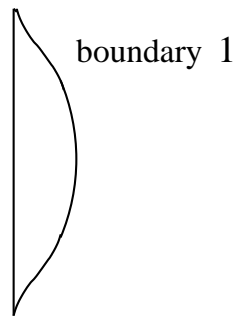}\\
\end{center}
\caption{A Carter-Penrose diagram
of the AdS geometry.}
\label{AdS}
\end{figure}
In \cite{Maldacena:2001kr},
in the context of the AdS-CFT correspondence,
Maldacena identified the
type-2 fields
as living\footnote{The word ``living" here is used
metaphorically, and
the role of type-2 fields in
the dual bulk geometry 
should be made more precise by
further study. 
This is one of the purposes of
this paper.
See also \cite{Herzog:2002pc} for a related study.}
on the boundary
behind the black hole horizon.\footnote{%
That the eternal black hole in AdS space 
is related to an entangled state
in the CFT was observed in
\cite{Balasubramanian:1998de,Horowitz:1998xk}.}
This gives a nice explanation
for the necessity of the introduction 
of type-2 fields
in the real time formulation 
of the finite temperature field theory.%
\footnote{A description of 
the thermodynamical nature of
black holes by thermo field dynamics
was studied in \cite{Israel:1976ur} in early days.
For further investigations after 
\cite{Maldacena:2001kr}
see \cite{Kraus:2002iv,Fidkowski:2003nf}.
For recent studies 
with more emphasis
on the real time formalism,
which is closer to the interest of this article,
see \cite{%
Herzog:2002pc,%
Hartnoll:2005ju}.}

However, there is a small puzzle here:
Below the Hawking-Page transition temperature,
the thermodynamically preferred configuration 
in canonical ensemble is 
the thermal AdS without a black hole
\cite{Hawking:1982dh}.
Since it is a finite temperature system,
the dual gauge theory
should still be described by
the real time formulation for the finite temperature.
Thus one must conclude that
below the Hawking-Page transition temperature,
the type-2 fields in the dual gauge theory
cannot correspond to the
degrees of freedom behind the black hole horizon,
since there is no black hole at all.
But then, what is the dual bulk description
for the type-2 fields in this case?
Since the AdS geometry corresponds to
the confined phase in gauge theory side,
confinement should change
the role of the type-2 fields in the bulk.
In this article,
I will show that this is indeed the case.
In the confined phase, the role of type-2 fields
in the gauge theory
is to make up {\em bulk} type-2 fields of 
closed string field theory
on AdS at finite temperature
in the real time formalism.
Since the bulk is also
at finite temperature,
it is very natural that
the bulk theory 
also has type-2 fields of its own.
The discussions will be
in the leading order in the $1/N$ expansion,
which corresponds to
the classical theory in the bulk.
Since the Carter-Penrose diagram
is based on classical gravity,%
\footnote{In this article the term 
``gravity" will be used with
possible $\alpha'$ corrections in mind.
In other words, I assume that gravity
has its origin in closed string theory.
I also expect, as in
\cite{Aharony:2003sx,Aharony:2005bq},
that the Hawking-Page transition
observed in the
Einstein-Hilbert action
(the lowest order in the $\alpha'$ expansion)
continuously extends to its
$\alpha'$ corrected version all down
to the string scale curvature regime,
which is dual to the weakly coupled gauge theory.}
this is sufficient
to explain 
the non-existence of the world
behind the horizon in this case. 

An outline of the organization of 
this article is as follows:
In section \ref{Pert}, I review
the derivation of Feynman rules 
in the real time formulation 
of field theories
at finite temperature.
Then, I give a crucial prescription
for incorporating
the effect of confined phase background
in this formalism.
In section \ref{Anal}, I study
Feynman diagrams in large $N$ gauge theories
in the confined phase and show a large class
of them vanish.
To illustrate the mechanism which selects
the non-vanishing diagrams,
a simple example is given
in Appendix \ref{A}.
In section \ref{Surv}, I argue that
the surviving Feynman diagrams
can be interpreted as
tree diagrams of
closed string field theory on AdS
at finite temperature in the real time formalism.
Section \ref{Summ} is devoted to
summary and discussions.

\section{Perturbative method in confined phase 
in the real time formalism}\label{Pert}

\subsection{Perturbative method in the real time formalism}

\begin{figure}
\begin{center}
 \leavevmode
 \epsfxsize=100mm
 \epsfbox{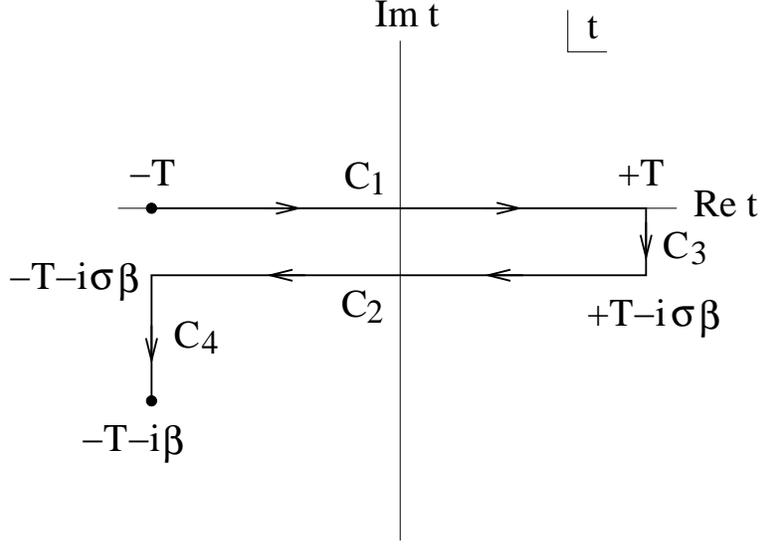}\\
\end{center}
\caption{The contour $C$}
\label{contour}
\end{figure}

In this section, I will review 
the perturbative method\footnote{One may
feel it slightly odd to study 
the confined phase 
by perturbative method, but as mentioned above,
the expectation here is that,
as in \cite{Aharony:2003sx,Aharony:2005bq},
the deconfinement
phase transition observed at weak coupling
persists continuously up to strong coupling.}
in
the real time formulation of finite temperature
field theories coupled to a gauge field.
A concrete example 
in mind
is the ${\cal N}=4$ super Yang-Mills
theory with $SU(N)$ gauge group on $S^3$
in the 't Hooft limit
\cite{Witten:1998qj,Witten:1998zw,%
Sundborg:1999ue,Polyakov:2001af,Aharony:2003sx},
but the method itself is applicable to 
general gauge theories in the 't Hooft limit.
To illustrate the point,
I take a real
scalar field $\Phi_{ab}(t)$
in the adjoint representation
of $SU(N)$ as an example.
Here, $a,b$ are $SU(N)$ gauge indices.
It is straightforward to include several
scalar fields, 
fermions or dynamical gauge fields.
Since I am interested in
the low temperature confined phase,
it is sufficient 
to study
quantum mechanics
obtained by dimensional
reduction 
of spatial coordinates on $S^3$.
(This approximation is valid 
when the inverse temperature $\beta$
and the length scale of interest
are much larger than the radius of the $S^3$.)
It is not difficult
to
generalize the discussion to more
general spatial manifolds, not necessarily
with dimensional reductions,
by replacing 
the mass in the discussions below to
the eigenvalues of the spatial Laplacian.
The quantities of interest 
are
thermal Green's functions
$G_{\beta}(t_1,\cdots,t_n)$
of the time-ordered products%
\footnote{Presisely speaking, these are 
the so-called
T$^*$-products since I
will eventually be interested
in the quantities obtained by path integral.}
of operators
$\hPhi(t)$ in Heisenberg picture:
\bea
 \label{TG}
G_{\beta}(t_1,\cdots,t_n)
=
\frac{1}{\Tr\, e^{-\beta \hH}}
\Tr
\left\{ e^{-\beta \hH}
T[\hPhi(t_1) \cdots \hPhi(t_n)]
\right\},
\eea
where ``$\Tr$" is the trace over 
{\em physical} 
states
satisfying the Gauss' law constraints:
\bea
 \label{phys}
\hat{\rho}_{ab}\, |phys \rangle = 0.
\eea
Here
$\hat{\rho}_{ab} = 
i ([\hPhi,\hPi_\Phi])_{ab}$ 
is the generator of
the gauge transformation, where
$\hPi_{\Phi ab}$ is the conjugate momentum
of $\hPhi_{ab}$.
$T[\cdots]$ denotes the time ordering
and $\beta$ is the inverse temperature.
The Hamiltonian $\hH$ is given by
\bea
 \hH =
\tr 
\left\{
\frac{g^2}{2}\hPi_{\Phi}\hPi_{\Phi}
+ \frac{\w^2}{2g^2} \hPhi^2 + \frac{1}{g^2} V[\hPhi]
\right\},
\eea
where 
$g$ is a gauge coupling constant,
``$\tr$" is a trace over the $SU(N)$ 
gauge group indices and
$V[\hPhi]$ is a potential term.
The mass $\w$ is proportional to 
the inverse radius of $S^3$ in the case when
the quantum mechanics
is obtained from the compactification 
of four dimensional conformal field theory 
on $S^3$.
In order to evaluate
the thermal Green's functions
by the path integral method,
the support of the field variables
should be extended
to the whole complex $t$-plane as in
\cite{Niemi:1983nf}:
\bea
\hPhi(t) = e^{i \hH t} \hPhi(0) e^{- i \hH t}.
\eea
One would like to have a 
functional representation for
the generating functional $Z[J]$ such that:
\bea
G_{\beta}(t_1,\cdots,t_n)
=
\frac{1}{Z(0)}
\frac{1}{i^n}
\frac{\delta}{\delta J(t_n)} \cdots
\frac{\delta}{\delta J(t_1)}
Z[J] \Biggr{|}_{J=0}  .
\eea
The functional
\bea
\Tr
\left\{
e^{-\beta \hH}
T[e^{i \int_{-T}^{T} dt J(t)\hPhi(t)}]
\right\}
\eea
has this property when $-T < t_i < T$.
But in order to calculate the thermal 
Green's function perturbatively, 
one should extend the $t$ integration
to a contour $C$ on the complex plane:
\bea
Z[J] 
= 
\Tr 
\left\{ 
e^{-\beta \hH}
T_C[e^{i \int_C dt J(t)\hPhi(t)}]
\right\}
\eea
where the contour $C$ is depicted in 
Fig.\ref{contour}.
$T_C[\cdots]$ denotes time-ordering
along the contour $C$.
By inserting a complete set of physical states
satisfying (\ref{phys}),
one arrives at the path integral representation
of the generating functional $Z[J]$:
\bea
 \label{PI}
Z[J] 
= 
\int [{\cal D}_C \Phi] \, 
e^{i \int_C dt 
\left\{
{\cal L}[\Phi] + J(t)\Phi(t)
\right\}},
\eea
where ${\cal L}[\Phi]$ is the Lagrangian%
\bea
{\cal L}[\Phi] = 
\frac{1}{g^2}
\tr 
\left\{ 
\frac{1}{2}D_t \Phi D_t \Phi
-\frac{\w^2}{2} \Phi^2 - V[\Phi]
\right\} .
\eea
The covariant derivative is given by
\bea
(D_t \Phi )_{ab} 
= 
\pa_t \Phi_{ab}
- i [A_0,\Phi]_{ab}.
\eea
The gauge field $A_0$ is introduced
while imposing the Gauss' law constraints
as delta function.
The path integral is over the fields
which satisfy the boundary conditions
\bea
 \label{bc}
\Phi_{ab}(-T-i\beta)
=
\Phi_{ab}(-T)  ,
\eea
following from the trace over 
the Hilbert space in (\ref{TG}).
One can rewrite (\ref{PI}) as
\bea
 \label{ZJ}
Z[J] 
= 
\exp
\left\{
{-i \int_C dt 
V\left[\frac{1}{i}\frac{\delta}{\delta J}\right]}
\right\}
\exp
\left\{
{-\frac{i}{2} g^2 \int_C dt \int_C dt' 
J_{ab}(t) D^C_{ab,cd}(t-t') J_{cd}(t')}
\right\},\nn
\eea
where the thermal propagator
$D^C_{ab,cd}(t-t')$
is a Green's function on the contour
\bea
 \label{green}
(- \pa_t^2 - \w^2) D^C_{ab,cd} (t-t')
=
\delta_C (t-t') \delta_{ad} \delta_{bc}
\eea
subject to the boundary condition
following from (\ref{bc}).
Here $\delta_C (t-t')$ is the delta function
defined on the contour \cite{Niemi:1983nf}:
\bea
\int_C dt \delta_C (t-t') f (t) = f(t').
\eea
The boundary time $T$  
is eventually taken to infinity.
By taking $J \rightarrow 0$ as $T \rightarrow \infty$,
the generating functional factorizes as
\bea
 \label{facto}
Z[J] = Z_{12}[J] Z_{34}[J] ,
\eea
where $Z_{12}[J]$ (respectively 
$Z_{34}[J]$) denotes the contribution
from the path $C_1$ and $C_2$ ($C_3$ and $C_4$).
The effect of finite temperature enters
in the propagators through the
boundary condition (\ref{bc}).
Although the generating functional can be
seen to factorize, $Z_{34}[J]$ part plays a role
for modifying the boundary conditions on
the Green's function, as I will explain below.

\subsection{Incorporating the effect of confined phase
 background}

In this subsection I present a prescription 
for reading off
the dual bulk description corresponding 
to the confined phase 
in the real time formalism.
When there are no external operator insertions,
one can take the Matsubara contour, i.e.
the line straight down from $-T$ to $-T-i\beta$.
Then the calculation reduces to that in the imaginary
time formalism,
where the confined phase is characterized 
by the vanishing
of the expectation value of the Polyakov loop.
The large $N$ saddle point value of the temporal
gauge field $A_0$ is given by
\bea
 \label{A0}
A_{0ab} 
= 
\delta_{ab} 
\frac{2\pi}{\beta N}
\left( a-\frac{N+1}{2} \right)
\qquad (constant)
\eea
in an appropriate gauge.\footnote{%
For weakly coupled 
gauge theories on $S^3$,
the low temperature phase
($\beta \gg \w^{-1}$)
is the confined phase 
\cite{Sundborg:1999ue,Polyakov:2001af,%
Aharony:2003sx,Aharony:2005bq}.
Above, I
wrote down
an action for a single scalar field
for which
there is no deconfinement transition
at zero 't Hooft coupling.
However, it is straightforward to
include several scalar fields
to have finite deconfinement temperature.
The saddle point (\ref{A0}) is evaluated
from the effective action
for $A_0$ obtained by integrating out
other massive fields,
which is justified in the
low temperature regime
\cite{
Aharony:2003sx,Aharony:2005bq}.}
This gives an appropriate expansion point
for perturbation theory
on the vertical parts of the contour.
In the imaginary time formalism, 
it was essential to
expand around the saddle point of $A_0$
(\ref{A0})
to read off the dual bulk geometry in the confined phase
\cite{Furuuchi:2005qm}.
I claim that
{\em also in the real time formalism,
the correct prescription
for reading off the dual bulk description
corresponding to the confined phase
is to include the saddle point value (\ref{A0})
of $A_0$ 
into the Green's function 
on the vertical parts of the contour}. 
Thus, instead of (\ref{green}),
I use Green's function which satisfies
\bea
 \label{boxA}
(-D_t^2 - \w^2) D^C_{ab,cd} (t-t')
=
\delta_C (t-t') \delta_{ad} \delta_{bc} \quad ,
\eea
where on the vertical parts of the contour
I have included the saddle point value of 
$A_0$ (\ref{A0}) 
in the covariant derivative $D_t$.\footnote{%
On the horizontal parts of the contour
one can choose $A_0 = 0$ gauge.}
Since I am including
the effect of the $A_0$ configuration (\ref{A0})
on the vertical parts of the contour, 
it is convenient to
define the field\footnote{%
The reason {Im} $t$, rather than
$-i t$ which is familiar from the case with
chemical potential, appears here is that
$A_0$ has non-zero expectation value
only on the vertical parts of the contour.}
\bea
 \label{redf}
\tilde{\Phi}_{ab}(t)
=
e^{-2\pi i\frac{a-b}{\beta N} (\mbox{\footnotesize Im}\, t)}
\Phi_{ab}(t)
\eea
so that the differential equation for the
Green's function $\tilde{D}^C(t-t')$ for $\tPhi(t)$
takes the form of the ordinary one (\ref{green}):
\bea
 \label{tbox}
(-\pa_t^2 - \w^2) \tilde{D}^C_{ab,cd} (t-t')
=
\delta_C (t-t') \delta_{ad} \delta_{bc}  \quad .
\eea
However, the field redefinition (\ref{redf})
modifies the boundary condition (\ref{bc}) to
\bea 
 \label{tbd}
\tilde{\Phi}_{ab}(-T-i\beta)
=
e^{2\pi i\frac{a-b}{N}}
\tilde{\Phi}_{ab}(-T) .
\eea
One can solve (\ref{green}) with the ansatz
\bea
\tilde{D}^C_{ab,cd} (t-t')=
\theta_C(t-t') \tilde{D}^>_{ab,cd} (t-t')
+
\theta_C(t'-t) \tilde{D}^<_{ab,cd} (t-t') ,
\eea
where $\theta_C(t-t')$ is the step function
defined on the contour \cite{Niemi:1983nf}:
\bea
\theta_C(t-t') 
= \int_C^t dt'' \delta_C(t''-t').
\eea
Since from
(\ref{PI}) to (\ref{ZJ})
the change of variable
\bea
\tPhi_{ab}(t) \rightarrow
\tPhi_{ab}(t) + \int_C dt' \tD_{ab,cd}(t-t') J_{cd}(t'),
\eea
has been made,
the boundary condition (\ref{tbd}) implies
\bea
 \label{bcD}
\tilde{D}^>_{ab,cd} (t-t'-i\beta)
=
e^{2\pi i\frac{a-b}{N}} \tilde{D}^<_{ab,cd} (t-t') .
\eea
The unique solution to (\ref{tbox}) 
with the boundary condition (\ref{bcD}) is
\bea
 \label{DABCD}
&&\tilde{D}^C(t-t')_{ab,cd}
=
\frac{-i}{2\w} 
\left[
(A e^{-i\w t} + B e^{i\w t}) \theta_C(t-t')
+
(C e^{-i\w t} + D e^{i\w t}) \theta_C(t'-t)
\right] \nn
\eea
with
\bea
 \label{ABCD}
A =
\frac{1}{1 - e^{-\beta \w - 2\pi i \frac{a-b}{N}}} ,
&&
B =
\frac{e^{-\beta \w + 2\pi i \frac{a-b}{N}}}%
{1 -  e^{-\beta \w + 2\pi i \frac{a-b}{N}}} , \nn
C =
\frac{e^{-\beta \w - 2\pi i \frac{a-b}{N}}}%
{1 -  e^{-\beta \w - 2\pi i \frac{a-b}{N}}}  ,
&&
D =
\frac{1}{1 - e^{-\beta \w + 2\pi i \frac{a-b}{N}}} .
\eea
The Green's function
(\ref{DABCD}) can be rewritten in the
spectral representation:
\bea
 \label{tDC}
i \tilde{D}^{C}_{ab,cd} (t-t')
=
\int_{-\infty}^\infty \frac{dk_0}{2\pi} 
e^{-i k_0 (t-t')}
\rho (k_0) 
[\theta_C(t-t')+N(k_0,a-b)]\delta_{ad} \delta_{bc}
\quad ,
\eea
where
\bea
\rho (k_0) = 2\pi \ve(k_0) \delta(k_0^2 -\w^2)
, \quad \ve(k_0) = \theta(k_0) - \theta(- k_0)
\eea
and
\bea
\label{Nk}
N(k_0,a-b) 
= \frac{1}{e^{\beta k_0 + 2\pi i \frac{a-b}{N}} - 1}.
\eea
As in (\ref{facto}), 
the partition function 
factorizes.
Therefore, only 
the propagators between the fields 
on the contours $C_1$ or $C_2$
need to be considered. 
The propagators for general $\sigma$
($0< \sigma < 1$, where 
$\sigma$ is given in Fig.\ref{contour})
are obtained as
\bea
\tD^{(11)}_{ab,cd}(t-t') 
&=& \tD^C_{ab,cd}(t-t'), \label{D11}\\
\tD^{(22)}_{ab,cd}(t-t')
&=& \tD^C_{ab,cd}((t-i\sigma\beta)-(t'-i\sigma\beta)), \label{D22}\\
\tD^{(12)}_{ab,cd}(t-t') 
&=& \tD^<_{ab,cd}(t-(t'-i\sigma\beta)), \label{D12}\\
\tD^{(21)}_{ab,cd}(t-t') 
&=& \tD^>_{ab,cd}((t-i\sigma\beta)-t')\label{D21}.
\eea
Notice that the propagator takes the form of
a $2\times 2$ matrix.
This means that the degrees of freedom 
are doubled compared with the original theory
(the theory at zero temperature)
\cite{Niemi:1983nf}.
The doubling of the degrees of freedom
originates from the two parts of the 
contour $C_1$ and $C_2$
in Fig.\ref{contour}.
$\tD^{(11)}$ (respectively $\tD^{(22)}$) is a propagator between 
type-1 (type-2) fields, and
$\tD^{(12)}$ and $\tD^{(21)}$ are mixed 
propagators
between type-1 and type-2 fields.


By taking $\sigma = \frac{1}{2}$
as in \cite{Niemi:1983nf,Niemi:1983ea},
one arrives at the most symmetric expression.
It is convenient to split the
propagator into a temperature dependent part 
and an independent part. 
Also, at this point it is convenient
to undo the field redefinition
(\ref{redf}) 
to obtain a symmetric expression 
for the propagators.\footnote{The 
phase appears symmetrically
in (1-2) and (2-1) components
of the propagator (\ref{Db}).}
In momentum space,
they are given by
\bea
 \label{dvD}
i D^{(rs)}_{ab,cd}
=
i D^{(rs)}_{0 ab,cd}  +
i D^{(rs)}_{\beta ab,cd}
\quad (r,s = 1,2),
\eea
\bea
 \label{D0}
i D_{0 ab,cd} =
\delta_{ad}\delta_{bc}
\left(
\begin{array}{cc}
 \frac{i}{k_0^2-\w^2+i\e} & 0 \\
 0 & \frac{-i}{k_0^2-\w^2-i\e}
\end{array}
\right) ,
\eea
\bea
 \label{Db}
i D_{\beta ab,cd} &=&
\delta_{ad}\delta_{bc}
\pi \delta(k_{0}^2- \w^2) \nn
&&
\times
\frac{1}{e^{{|\beta k_0+2\pi i\frac{a-b}{N}}|_R}-1} 
\left(
\begin{array}{cc}
 1 & e^{\frac{1}{2}|\beta k_{0}+2\pi i\frac{a-b}{N}|_R} \\
e^{\frac{1}{2}|\beta k_{0}+2\pi i\frac{a-b}{N}|_R} & 1
\end{array}
\right) .
\eea
In the above, $|\cdots |_R$ is defined as
\bea
 \label{||R}
|z|_R = 
\left\{
\begin{array}{c}
z \quad (\mbox{Re}\, z > 0)\\
-z \quad (\mbox{Re}\, z < 0)
\end{array}
\right.  , \quad (\mbox{Re}\, z \ne 0).
\eea
Eq.(\ref{||R}) is not defined for
$\mbox{Re}\, z = 0$, but 
because
of the on-shell delta function in
(\ref{Db}) one does not need to consider
that case as long as $\w \ne 0$.\footnote{In the case
which the quantum mechanics 
is obtained from the compactification
of four dimensional conformal field theory 
on $S^3$,
$\w$ is proportional to 
the inverse radius of $S^3$ and non-zero.}
\begin{figure}
\begin{center}
 \leavevmode
 \epsfxsize=100mm
 \epsfbox{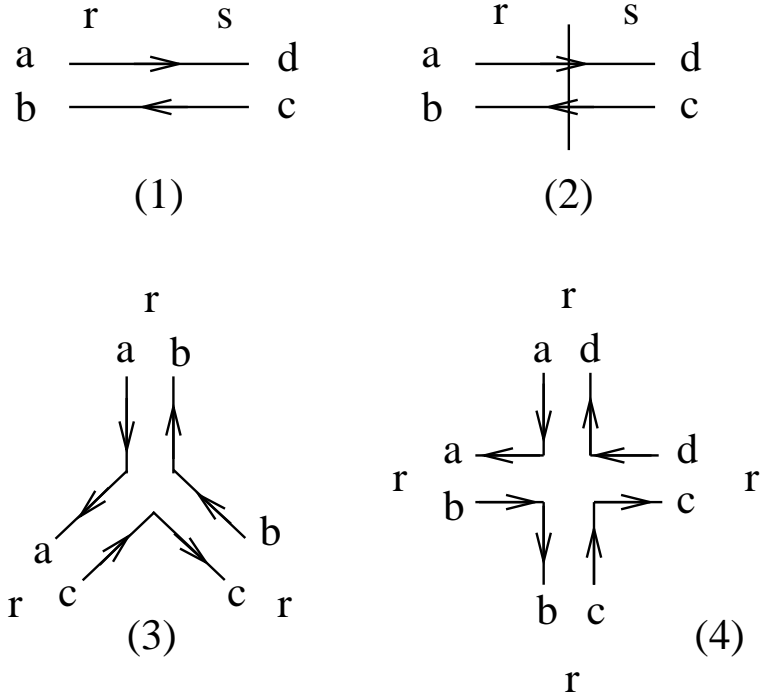}\\
\end{center}
\caption{Feynman rules for the real time formulation of
large $N$ gauge theories at finite temperature:
(1) is the temperature independent part
of the propagator (\ref{D0}), while
(2) is the temperature dependent part
(\ref{Db}).
The index flow indicated by the arrow
is directed from 
the first matrix index to the second matrix index
of adjoint fields.
The temperature dependent part
of the propagator (2) is drawn with
a ``cut" line (the vertical line in the figure).
The interaction vertices (3),(4) are drawn
schematically, just to show the index flow structure.
One can assign a more detailed structure
to the interaction vertices
according to the action of one's interest.
The interaction vertices do not mix the type-1 fields
with the type-2 fields.}
\label{Feyn}
\end{figure}

Perturbative Feynman rules can be obtained 
just as in the conventional field theories
and are sketched in Fig.\ref{Feyn}.
I have adopted the 't Hooft's double-line
notation \cite{'tHooft:1973jz}.
(1) represents the temperature independent
part of the propagator $iD^{(rs)}_{0 ab,cd}$
and (2) 
the temperature
dependent part 
$iD^{(rs)}_{\beta ab,cd}$.
The temperature dependent part
of the propagator is drawn with
the ``cut" line in Fig.\ref{Feyn} (2).
{\em This cut is one of the most important 
tools I will use repeatedly 
in the following discussions.}
Type-1 fields and type-2 fields
are coupled only through the propagators:
The interaction vertices do not mix
type-1 and type-2 fields.
The interaction vertices of type-2 fields
are given by the complex conjugate
of those of type-1 fields:
\bea
 \label{rp}
i\, \tr V_2[\Phi_{(2)}] 
= 
\left.(i\, \tr V_1[\Phi_{(1)}])^*
\right|_{\Phi_{(1)}\rightarrow \Phi_{(2)}}.
\eea
I have assumed that the potential is real.

\section{Analysis of Feynman 
diagrams in the confined phase}\label{Anal}

In this section,
with the prescription for incorporating
the effect of 
the confined phase background (\ref{A0})
discussed in the previous section,
I will show that
contributions from a large class of 
Feynman diagrams vanish.
In Appendix \ref{A}, a simple example
of the following discussions is provided.
The reader may find it helpful 
to read them in parallel.

The quantities of interest 
in this article
are the correlation functions
of gauge invariant 
single trace local operators,
which correspond to closed string states
in the AdS-CFT correspondence.
Throughout this article I will work
in the planar limit $g \rightarrow 0$,
$N \rightarrow \infty$ with
the 't Hooft coupling $g^2 N$ fixed.
In the planar limit,
one can always associate
a loop momentum to an index loop
\cite{Furuuchi:2005qm,Furuuchi:2005eu}, 
as will be explained below.%
\footnote{There are $\ell + 1$ index loops for
$\ell$ (momentum) loop planar diagrams
of a correlation function of
gauge invariant operators,
but one summation over gauge indices
decouples since the gauge indices always
appear as a difference of two indices
\cite{Furuuchi:2005qm,Furuuchi:2005eu}.}
For a given 't Hooft-Feynman diagram,
I draw
a tree sub-diagram 
which connects the external legs,
to show the flow of the external momentum
(Fig.\ref{vanish}), between the double lines.
The total momentum on a propagator
is a sum of two momenta
associated with the index lines
(taking into account the sign indicated by the arrows),
and an external momentum flow
if there is any.
By a shift of loop momenta, which are 
integration variables,
one can choose any tree sub-diagram
connecting the external legs
to express the external momentum flow.
But once it is chosen,
the integrations over loop momenta
should be done
with
that fixed external momentum flow.

\subsection{Diagrams which have an index loop 
with only one cut -- vanish}

I first consider Feynman diagrams which have
at least one index loop containing
only one cut (which is denoted as $a_i$ below).
In this case the cut must be either
1-1 or 2-2 cut.\footnote{When there is a 1-2 cut
on an index loop, there must be a 2-1 cut
on that index loop. See the explanation
at the end of subsection \ref{somesurvive}}
From (\ref{Db}),
the diagram with the 1-1 or 2-2 cut
is proportional to a factor
\bea
 \label{phase}
\sum_{a_i=1}^N 
\frac{1}{e^{|\beta(p_{0i}-p_{0j})
+2\pi i \frac{a_i-a_j}{N}|_R}-1}.
\eea
Here, $i,j$ label the index-momentum loops.
As mentioned earlier, 
the loop momentum $p_{0i}$
is ``associated" with the gauge index $a_i$,
that means, they always appear
in the combination 
$\beta p_{0i}
+2\pi i \frac{a_i}{N}$.
The origin of this combination 
is the covariant derivative for 
adjoint fields.
Therefore, by taking the background gauge
around the field configuration (\ref{A0}),
even when there are derivative couplings
the loop momentum and the associated index
also appear in the same combination.
Such derivative couplings give rise
to a multiplicative factor 
which is polynomial in 
$\beta p_{0i}+2\pi i \frac{a_i}{N}$.
Those just require a minor modification
in the following discussions
and do not change the conclusion about
whether a diagram vanishes or not.
Therefore, to keep the essential points clear 
in the presentation,
I will only write down the formula 
for the case in which such
derivative couplings are not involved.
Since I am working in the strict $N \rightarrow \infty$
limit, 
the sum over 
the gauge indices
$a_i$ can be replaced by integral:
$\frac{a_i}{N} \rightarrow \theta_i$,
$\sum_{a_i=1}^N \rightarrow N \int_0^1 d\theta_i$.
(To avoid repetition,
this replacement will be
implicit in what follows.)
Then, one can Fourier expand
the integrand as
\bea
\label{vmec}
\int_0^1 d\theta_i 
\frac{1}{e^{|\beta(p_{0i}-p_{0j})
+2\pi i (\theta_i-\theta_j)|_R}-1}
&=&
\int_0^1 d\theta_i 
\sum_{n=1}^\infty 
e^{-n |\beta (p_{0i}-p_{0j})
+2\pi i (\theta_i-\theta_j)|_R} \nn
&=& 0 .
\eea
From the definition (\ref{||R}), 
this kind of diagram
has either {all} negative (when $p_{0i}>p_{0j}$)
or {all} positive (when $p_{0i}<p_{0j}$)
powers of $e^{2\pi i \theta_i}$.
In either case,
(\ref{vmec}) vanishes.\footnote{%
One does not need to worry about the case where
$n$ is a multiple of $N$ in 
(\ref{vmec})
in the strict
$N \rightarrow \infty$ limit.
Since the $N \rightarrow \infty$ limit is taken
before the Fourier expansion.
I thank S. Kalyana Rama and A. Sen for
questions and comments on this point.}
{\em Eq.(\ref{vmec}) is the basic equation
relevant for 
selecting non-vanishing Feynman diagrams
in the confined phase,
and will repeatedly appear in the following}.

\subsection{The case in which
a cut-out loop divides
a diagram into two disconnected pieces,
one of which does
not contain the external legs 
-- still vanish}\label{stillvanish}

\begin{figure}
\begin{center}
 \leavevmode
 \epsfxsize=70mm
 \epsfbox{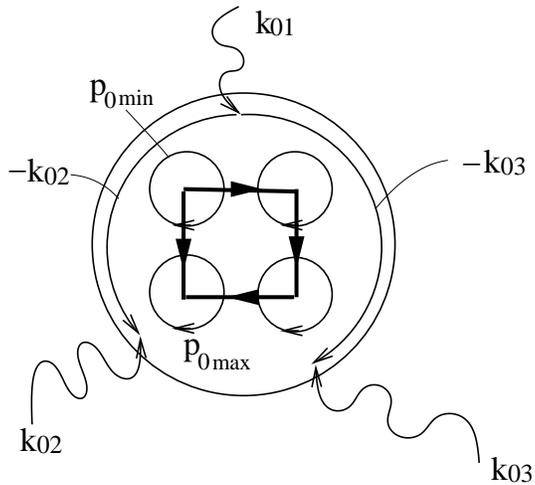}\\
\end{center}
\caption{A diagram in which
the sequence of the bold arrows surrounds
a region without the external legs.
The starting
and the end points of the
bold arrows are connected
when they are inside of the same index loop.
The bold arrows on the cuts are directed from
the smaller
to the larger momenta, 
following the rule given in Fig.\ref{boldarrow}.
In this case, there are 
a maximum and a minimum loop momenta
among the momenta flowing through the cuts.
The maximum momentum loop
is the one where all the bold arrows come in,
and the minimum momentum loop
is the one where all those go out.
Such diagrams vanish
after the summation over the gauge 
index $a_{max}$ or $a_{min}$.}
\label{vanish}
\end{figure}

In the previous subsection,
I have shown that
the diagrams which have an index loop
with only one cut vanish.
Therefore, below 
I will consider the cases
in which all the index
loops either contain no cut or more than
one cut.
In these cases,
if there is a cut on an index loop,
there must be at least one more cut
on this index loop.
Since 
the sequence of cuts
cannot end on an index loop,
they make up a closed circuit
(when one connects the end points of the cuts
which are inside the same index loop)
which
cut the diagram into disconnected pieces.
In this subsection,
I study the case in which
a sequence of the cuts make up a loop,
and this ``cut-out loop"
divides a diagram
into two disconnected pieces,
one of which does not contain
the external legs (Fig.\ref{vanish}).
\begin{figure}
\begin{center}
 \leavevmode
 \epsfxsize=70mm
 \epsfbox{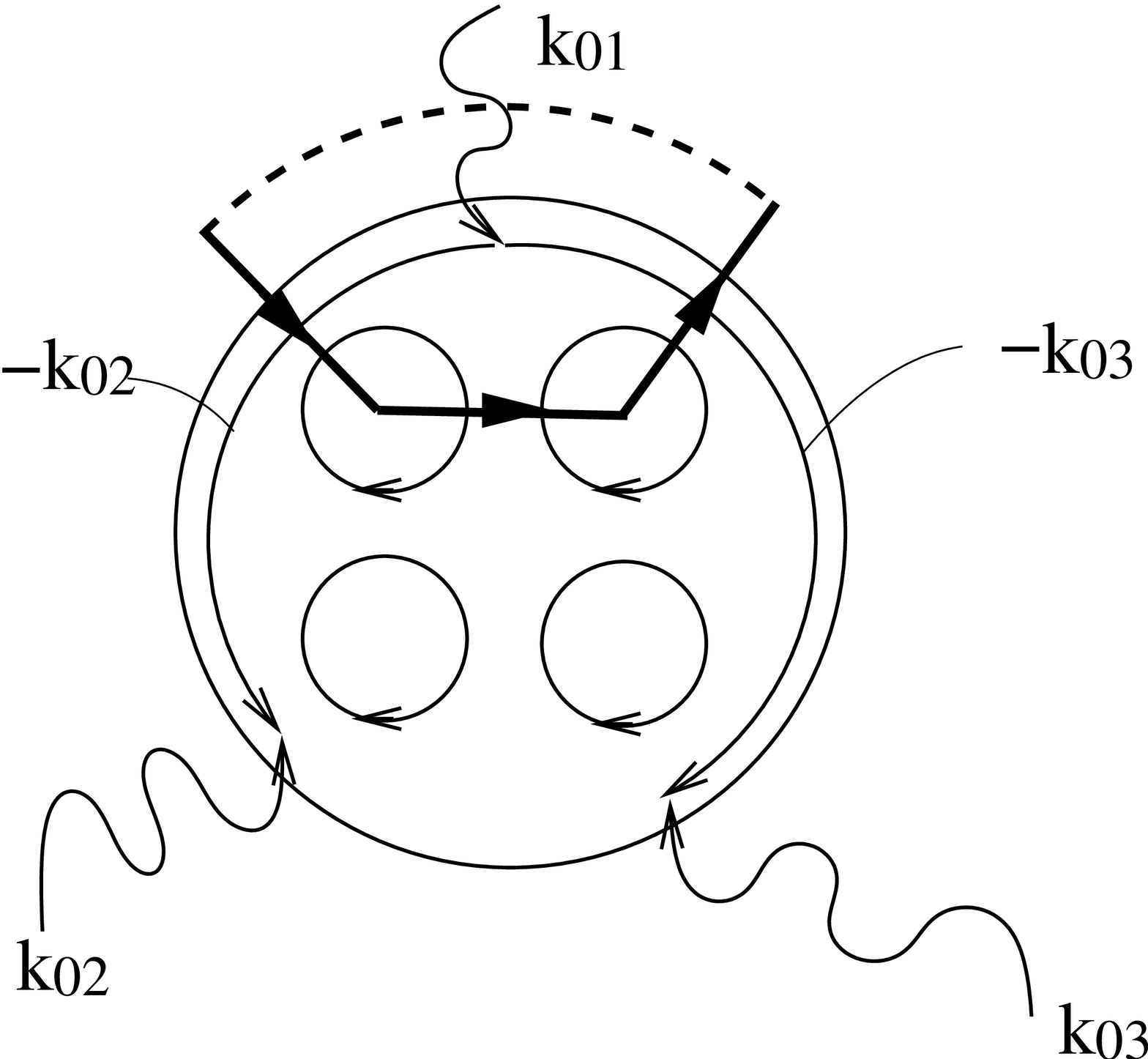}\\
\end{center}
\caption{%
In this figure, 
the starting and the end points
of the sequence of the bold arrows is
connected, since
they are inside 
of the same index loop (planar
diagrams can be thought of 
as drawn on a sphere), 
as indicated
by the dashed line.
Thus the sequence of the bold arrows
make up a cut-out loop.
In non-vanishing diagrams,
all cut-out loops divide 
the diagram into two disconnected pieces,
both containing 
the external legs. 
In the above figure,
the region containing
the external leg with momentum $k_{01}$
is separated from the region 
containing the external legs
with momentum $k_{02}$ and $k_{03}$
by the cut-out loop.}
\label{cutout}
\end{figure}
\begin{figure}
\begin{center}
 \leavevmode
 \epsfxsize=80mm
 \epsfbox{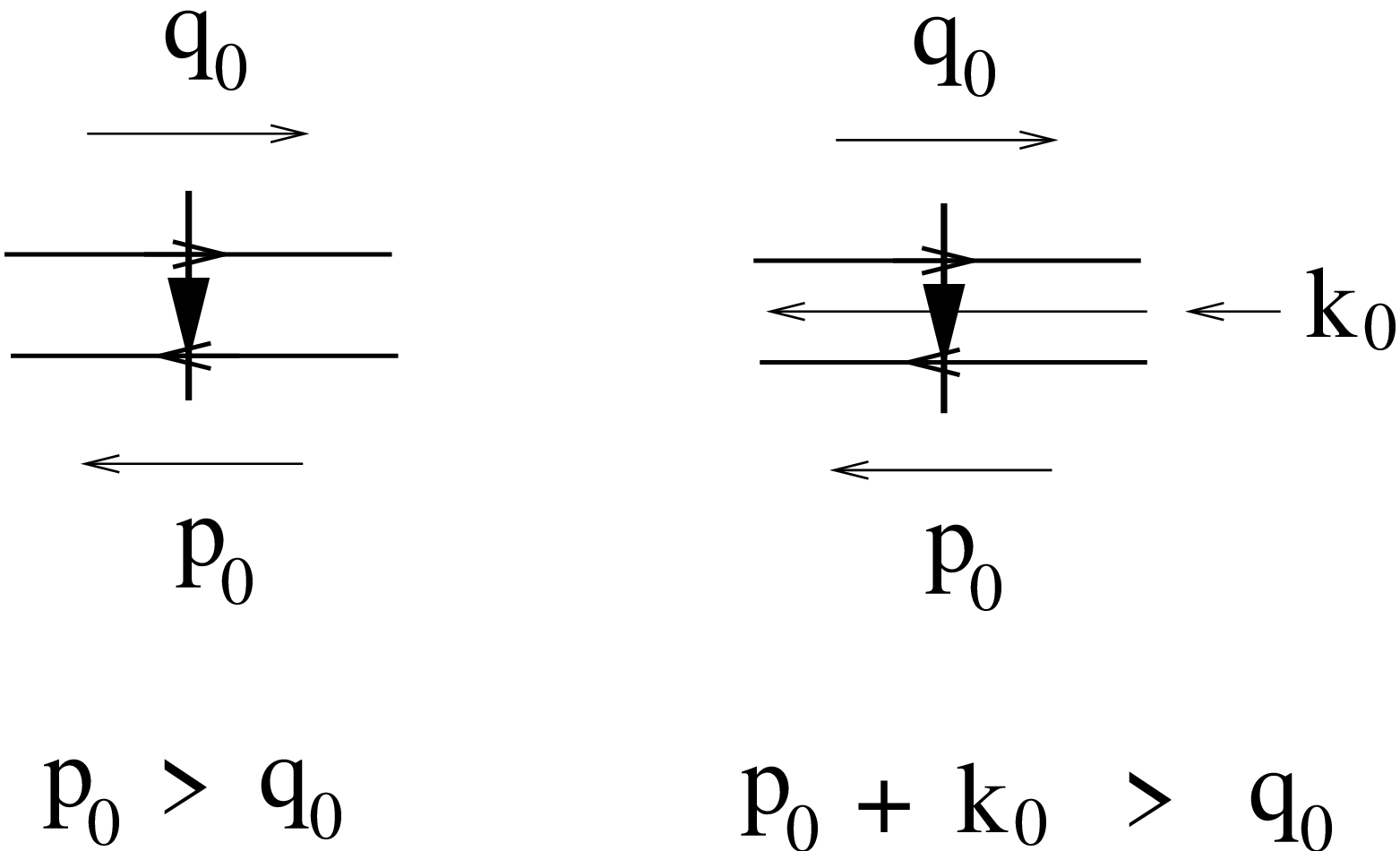}\\
\end{center}
\caption{A bold arrow is put on the cut
(temperature dependent part of the propagator,
see Fig.\ref{Feyn}) to indicate
which of the momenta associated
with the index lines is larger. 
The direction is from the smaller
to the larger momenta, and
the sign of the loop 
momenta are determined referring to the
direction of the associated index lines.
When there is an external momentum 
flowing into a propagator 
($k_0$ in the right),
it is added to the momentum with the same direction.
In other words, the direction of the bold arrow
is determined according to the 
direction of the total momentum flowing on
the propagator.}
\label{boldarrow}
\end{figure}
In this case,
one can put the external momentum flow
avoiding the region cut out by the cut-out loop.
Without loss of generality,
one can assume that there are no further
cut-out loops inside\footnote{%
For a loop on a sphere topology,
inside and outside
is a relative notion,
but one may call one side inside
and the other outside.
Here I called the region
of the Feynman diagram
which does not contain the
external legs as inside.}
the one under consideration.
Then, since the type-1 and type-2 fields
only mix through 1-2 or 2-1 cut propagators,
the cut out region consists of 
either entirely
type-1 or entirely type-2 propagators and vertices.
I first study the case in which
the cut-out loop is either
made of entirely 1-1 cuts or entirely 2-2 cuts.
Other cases can be treated similarly.
The correlation function is proportional to
a product of the factors
coming from the 1-1 or 2-2 cuts on
this cut-out loop (see (\ref{Db})):
\bea
 \label{expand}
\frac{1}{e^{|\beta (p_{0i}-p_{0j})
+2\pi i \frac{a_i-a_j}{N}|_R}-1}
=
\sum_{n=1}^{\infty}
e^{-n  |\beta(p_{0i}-p_{0j})+
2\pi i\frac{a_i-a_j}{N}|_R}  .
\eea
Contrary to the previous case, since
$|\beta (p_{0i}-p_{0j})+2\pi i\frac{a_i-a_j}{N}|_R$
may have different
signs for different combinations of
$(i,j)$,
the phases may cancel 
in the product.
Then the summation over these
gauge indices
can have a non-zero value.
However, 
there is always a maximum loop momentum, 
$p_{0max}$,
and a minimum
loop momentum, $p_{0min}$,
among
the loop momenta which cross the cuts
on the cut-out loop.
I denote the associated gauge indices
$a_{max}$ and $a_{min}$, respectively.
Then, from the definition (\ref{||R}), 
\bea
 \label{|max|R}
\left|\beta(p_{0max}-p_{0i})+2\pi i\frac{a_{max}-a_i}{N}\right|_R
= \beta(p_{0max}-p_{0i})+2\pi i\frac{a_{max}-a_i}{N}
\eea
and
\bea
 \label{|min|R}
\left|\beta(p_{0min}-p_{0i})+2\pi i\frac{a_{min}-a_i}{N}\right|_R
= 
- \left(
\beta(p_{0min}-p_{0i})+2\pi i\frac{a_{min}-a_i}{N}
\right),
\eea
i.e., they
have a definite sign 
in any combination with other $p_{0i}$.
(Recall that because of the on-shell delta function
in (\ref{Db}),
the difference between two momentum flows 
in a cut-propagator is always non-zero: 
$|p_{0i}-p_{0j}|= \w$.)
This means that
in the Fourier expansion (\ref{expand}),
the sign of the phase in
$e^{2\pi i\frac{a_{max}}{N}}$ 
{\em always} 
appears negative 
while that of 
$e^{2\pi i\frac{a_{min}}{N}}$ 
{\em always}
positive,
and hence do not cancel.
After summing over the index $a_{max}$ 
(or $a_{min}$),
the contribution from such diagrams
vanish, as in (\ref{vmec}).
In Fig.\ref{vanish}
the cut is drawn with a bold arrow 
which is directed 
from the smaller 
to the larger loop momenta,
following the rule given in Fig.\ref{boldarrow}.
The maximum loop momentum
$p_{0max}$ 
(among the loop momenta which cross
the cut-out loop)
is the loop
to which those bold arrows 
all come in,
and the minimum 
$p_{0min}$ is 
from which all those bold arrows flow out.

Essentially the same arguments
hold for 1-2 (2-1) cuts. 
In this case, for $p_{0max}$,
for example,
there is an extra overall factor
$e^{\frac{1}{2}(\beta(p_{0max}-p_{0i})
+2\pi i \frac{a_{max}-a_i}{N})}$
compared with the 1-1 or 2-2 cuts.
Since type-1 and type-2 fields
mix only through 1-2 or 2-1 cut propagators,
and there are no further cuts inside
the cut-out loop,
the cut out region is made of either
entirely type-1 fields or entirely type-2 fields.
Then, an index line
which enters from the type-1 region to the 
type-2 region must come out again.\footnote{%
Recall that every index loop has either 
zero or more than
one cut in order for a diagram not to vanish.}
Therefore, there are always an even number of 
the extra phase factors coming from the 1-2 (2-1) cuts
for each index loop, and they multiply up to
integer powers of 
$e^{\beta p_{0max} + 2\pi i \frac{a_{max}}{N}}$.
It partially cancels the relevant phase coming from
the Fourier expansion
of 
$1/(e^{ (\beta (p_{0max}-p_{0i})
+2\pi i\frac{a_{max}-a_i}{N})}-1)$,
but
it does not completely cancel
the phase factors. (This is 
basically because originally 
each
of these was just 
half of the relevant phase).

\subsection{The case in which 
every 
cut-out loop divides a diagram
into two disconnected pieces
both containing the external legs 
-- some diagrams survive}\label{somesurvive}

From the discussions in the
previous subsections, 
the non-vanishing contributions
arise only from diagrams
in which 
every cut-out loop
divides the diagram into two disconnected pieces
both 
containing the external legs
(Fig.\ref{cutout}).
Here, the closed circuits made of cuts
are interpreted as overlapping 
cut-out loops,
as will be explained in more detail
in the next subsection.
In this case,
the tree sub-diagram expressing
the external momentum flow
must cross the cut:
Since the diagram is completely
disconnected by the
cut-out loop, there is no way to 
connect the separated external legs
avoiding the cuts.
For this kind of diagrams, the earlier
argument does not hold, because
an additional external momentum
comes into the argument:
Now it is possible to have
$p_{0max} < p_{0i} + k_0$
or
$p_{0min} + k_0 > p_{0i}$
for some $i$,
where $k_0$ is an external momentum
flowing into the cut propagator
with index lines $a_{max}$ and $i$,
or $a_{min}$ and $i$, respectively.
From the definition (\ref{||R}),
this means
\bea
\left|\beta(p_{0max}-(p_{0i}+ k_0))+2\pi i\frac{a_{max}-a_i}{N}\right|_R
= - \left(\beta(p_{0max}-(p_{0i}+ k_0))+2\pi i\frac{a_{max}-a_i}{N}\right) 
\eea
for some $i$, or
\bea
\left|\beta(p_{0min}+ k_0-p_{0i})+2\pi i\frac{a_{min}-a_i}{N}\right|_R
= \beta(p_{0min}+ k_0-p_{0i})+2\pi i\frac{a_{min}-a_i}{N}
\eea
for some $i$. Compare these
with the previous cases
(\ref{|max|R}) and (\ref{|min|R}).
Still, in order for a diagram not to vanish,
there must not be an index loop
where bold arrows are all coming in
(as in the index loop associated with $p_{0max}$
in the previous case),
or all going out 
(as in that associated with $p_{0min}$
in the previous case).
In other words,
when there is a bold arrow
coming into an index loop,
there must be at least one 
bold arrow which comes out of it.
Thus the {\em directed} sequences of bold arrows
make up closed circuits.
Note that this can only happen
when the sequence of the bold arrows
crosses the external momentum flow,
otherwise it is inconsistent with
the definition of the bold arrows
(i.e., the directed sequence of bold arrows should 
be along an increasing sequence of momenta,
see Fig.\ref{boldarrow}).
In the next section,
these non-vanishing diagrams will be
studied in more detail.

\section{Surviving diagrams
as closed string tree diagrams
in the real time formulation in the bulk}\label{Surv}

Now I will argue that the 
non-vanishing Feynman diagrams
in the confined phase
can be interpreted
as tree diagrams
of the real time formulation of 
closed string field theory on AdS
at finite temperature.%
\footnote{%
The reason I am using the phrase
closed string {\em field} theory here is
just to indicate that there will 
be propagators
and interaction vertices for
the type-2 {\em fields} in the bulk.
The discussions remain within
perturbation theory.
Non-perturbative studies
using the field theory will be interesting
but beyond the scope of this article.}
Here, the word ``closed string theory" is used
in a loose sense, in that I regard
random surfaces
obtained from the Feynman diagrams of
the large $N$ gauge theories
as closed string worldsheets \cite{'tHooft:1973jz},
hoping more precise description
as a closed string theory will emerge from the
conjectured duality between closed string theories on
AdS 
and large $N$ gauge theories 
\cite{Maldacena:1997re}.%
\footnote{See 
\cite{Gopakumar:2003ns,Gopakumar:2004qb,%
Gopakumar:2005fx,Gopakumar:2004ys,Furuuchi:2005qm} 
for a recent attempt to
describe precisely how the large $N$ gauge theory
correlation functions organize themselves
into closed string amplitudes.}

\subsection{Cut-out loops as cuts of closed 
string propagators}

First,
I argue that 
the cut-out loops
can be identified with 
the cuts of the temperature dependent parts
of the closed string propagators.
In closed string field theory,
there is a freedom in
separating the propagator and
the interaction vertices 
(see e.g. \cite{Hata:1993gf}).
However, it
must be constructed
in a way that it reproduces
the correct perturbative diagrams.
Therefore, in the following
I do not
specify
how to divide the worldsheet into
the propagators
and interaction vertices,
but the cut-out loop must be within 
the propagator part.

I will show that those cut-out loops
give the correct
energy dependence
as the temperature dependent parts
of closed string propagators
in the real time formalism.
I will discuss this in the free field limit, i.e.
zero 't Hooft coupling limit.
I expect that the discussions 
can be extended to
finite 't Hooft coupling
if the full propagators
in the gauge theory
instead of the free propagators are used.%
\footnote{%
I thank R. Gopakumar for reminding me
that the following discussions
on the energy dependence
are for free field theory,
and suggesting 
how the generalization to
the finite 't Hooft coupling will be.}

\subsubsection{An isolated cut-out loop}

\begin{figure}
\begin{center}
 \leavevmode
 \epsfxsize=85mm
 \epsfbox{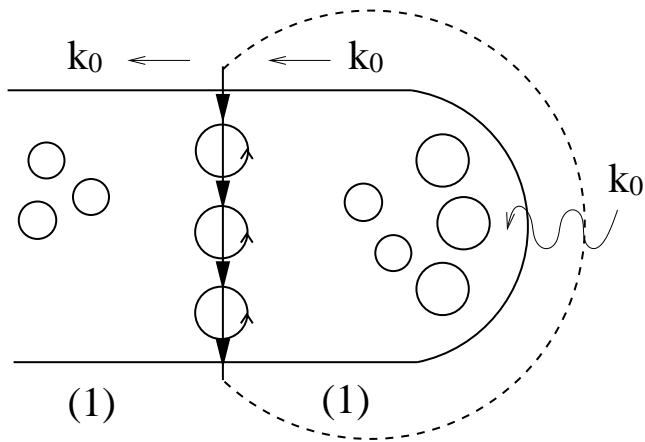}\\
\end{center}
\caption{A 1-1 cut-out loop
(the starting and the end points of
the sequence of the bold arrows
should be regarded as the same point,
as indicated by the dashed line).
The bold arrows are all in the same direction
on the cut-out loop
in order for the diagram not to vanish.
The figure can be regarded as the case
$k_0 = J\w= 4\w$, which is enforced
by the on-shell delta functions of the cuts
on the cut-out loop.
``(1)" in the above
refers to the type-1 region.}
\label{cut}
\end{figure}

\begin{figure}
\begin{center}
 \leavevmode
 \epsfxsize=70mm
 \epsfbox{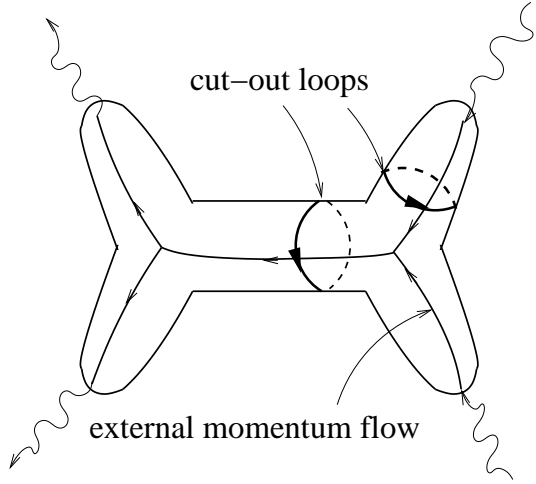}\\
\end{center}
\caption{A `t Hooft Feynman diagram
identified with a closed string worldsheet.
The cut-out loop must cross the tree sub-diagram
where the external momentum flows.
Thus the cut-out loop can be identified
with the cut on a closed string propagator.}
\label{tree}
\end{figure}

I start with
the case when
the cut-out loop is isolated,
i.e.
all the index loops crossed
by the cut-out loop 
have one incoming and one out going
bold arrow.
The case for more general closed circuits
will be explained shortly.
Notice first that the cut-out loop
is proportional to $\delta (k_0^2 - (J\w)^2 )$,
where $k_0$ is the momentum
flowing into the cut-out loop
and $J$ is 
the
number of the gauge theory propagators
cut by the cut-out loop (Fig.\ref{cut}).
This is due to the fact that 
on the cut-out loop,
the bold arrows must be directed in the
same direction,
in order for
the diagram not to vanish,
as described in the previous section.
This means that
the momenta flowing through those 
cuts are either all $\w$
or all $-\w$.
These give a factor
$\delta(k_0-J\w)$ or
$\delta(k_0 + J\w)$, respectively.
Up to these delta functions,
both contributions are the same
and one obtains $\delta (k_0^2 - (J\w)^2 )$.
$J\w$ corresponds
to the external momentum flow on
an edge of the tree sub-diagram
which the cut-out loop crosses,
and thus the cut-out loop
can be assigned to
a cut on an edge of the closed string tree
diagram, i.e., a closed string propagator
(Fig.\ref{tree}).
In the free field limit,
$J\w$ can be identified with 
the energy of a closed string state
corresponding to
$\tr \Phi^J$ 
\cite{Witten:1998qj,Gubser:1998bc}.\footnote{%
The concrete example in mind
is ${\cal N}=4$ super Yang-Mills theory
compactified on $S^3$,
and $\w$ is proportional to the 
inverse radius of the $S^3$.
Since  
this theory
is conformal, after a suitable rescaling 
only the ratio
of the two scales $\beta \w$
is physically relevant.}
Therefore, this is the correct 
on-shell delta function which should appear in
the temperature dependent part of 
the propagators in the real time formalism:
The temperature dependent
part of the propagator in the real time formalism
in general has a form of (\ref{Db})
without the gauge index dependent phase factors
specific to the gauge theory in the confined phase.
(Recall the derivation of (\ref{Db})
or see \cite{Landsman:1986uw,Leblanc:1987zj}.)

Furthermore,
the 1-1 and 2-2 cut-out loops
give rise to a following
gauge index summation:
\bea
 \label{cc11}
\sum_{a_1=1}^N \cdots \sum_{a_J=1}^N
\prod_{i=1}^{J}
\frac{1}%
{e^{\beta \w \pm 2\pi i \frac{a_i-a_{i+1}}{N}} -1}
&=&
\sum_{a_1=1}^N \cdots \sum_{a_J=1}^N
\prod_{i=1}^{J}
\sum_{n_i=1}^\infty
e^{-n_i(\beta \w \pm 2\pi i \frac{a_i-a_{i+1}}{N})} \nn
&=&
\sum_{n=1}^{\infty} e^{- n\beta \w J}\nn
&=&
\frac{1}{e^{\beta \w J} -1}  .
\eea
Here, $a_{J+1} = a_1$.
Recall that the total momentum of the cuts
on the cut-out loop are either all $\w$
or all $-\w$, and 
$\pm$ respectively
correspond to each case:
The contribution from
$\delta(p_{0i}-p_{0j}(+k_0) \mp \w)$,
where $p_{0i}$ and $p_{0j}$ are loop
momenta associated with the index 
$a_i$ and $a_j$,
and $k_0$ is a possible external momentum flow
(on one of the cuts).
Notice that 
from the first line to the second line, 
the sum over the gauge indices 
only picked up the
$n_1 = n_2 = \cdots = n_J \equiv n$ contributions,
since the phases must cancel to give a
non-zero result.
Similarly,
the 1-2 and 2-1 cut-out loops give rise to a factor
\bea
 \label{cc12}
\sum_{a_1=1}^N \cdots \sum_{a_J=1}^N
\prod_{i=1}^{J} 
\frac{%
e^{ \frac{1}{2}(\beta \w + 2\pi i %
\frac{a_i-a_{i+1}}{N})} }%
{e^{\beta \w \pm 2\pi i \frac{a_i-a_{i+1}}{N}} -1}
&=&
e^{\frac{\beta}{2}\w J}
\sum_{a_1=1}^N \cdots \sum_{a_J=1}^N
\prod_{i=1}^{J}
\sum_{n_i=1}^{\infty}
e^{- n_i (\beta \w \pm 2\pi i \frac{a_i-a_{i+1}}{N})} \nn
&=&
e^{\frac{\beta}{2}\w J}
\sum_{n=1}^{\infty} e^{- n \beta \w J}\nn
&=&
\frac{e^{\frac{\beta}{2}\w J}}{e^{\beta \w J} -1}.
\eea
Again, the factors
(\ref{cc11}) and (\ref{cc12})
are the expected ones for 
the 
temperature dependent parts of the
propagators
in the real time formalism.
All together, these give
the correct energy dependent 
factors to be interpreted
as the temperature dependent part of
the propagators
in the real time formulation of
closed string field theory at finite temperature.

\subsubsection{Closed circuits as 
overlapping cut-out loops}

\begin{figure}
\begin{center}
 \leavevmode
 \epsfxsize=70mm
 \epsfbox{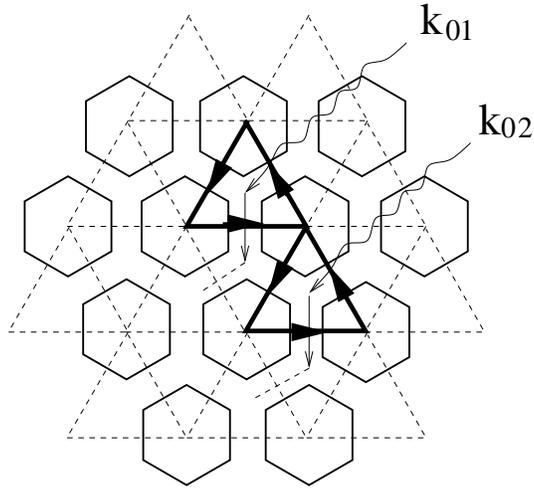}\\
\end{center}
\caption{A part of a planar 
't Hooft-Feynman diagram (the double lines)
and its dual graph (the dashed lines).
The cut-out lines (bold arrows)
make up a closed circuit
on the dual graph.
Each sub-loop of the closed circuits 
must 
surround a region with
the external leg, 
and there must be
at least one external leg on the other side,
as described in section \ref{somesurvive}.}
\label{circuit}
\end{figure}

\begin{figure}
\begin{center}
 \leavevmode
 \epsfxsize=70mm
 \epsfbox{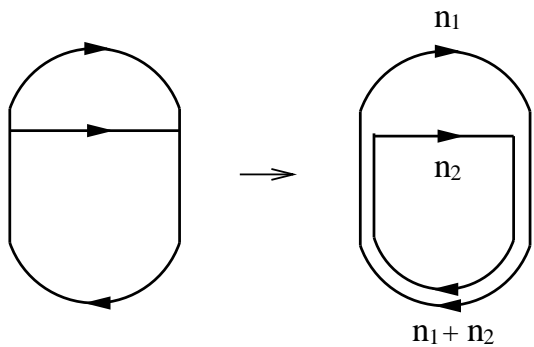}\\
\end{center}
\caption{A closed circuit
with the ``momentum conservation" at
each vertex.
The ``momentum conservations"
can be solved by assigning ``loop momenta"
to the sub-loops in the circuit.}
\label{circuit1}
\end{figure}

\begin{figure}
\begin{center}
 \leavevmode
 \epsfxsize=70mm
 \epsfbox{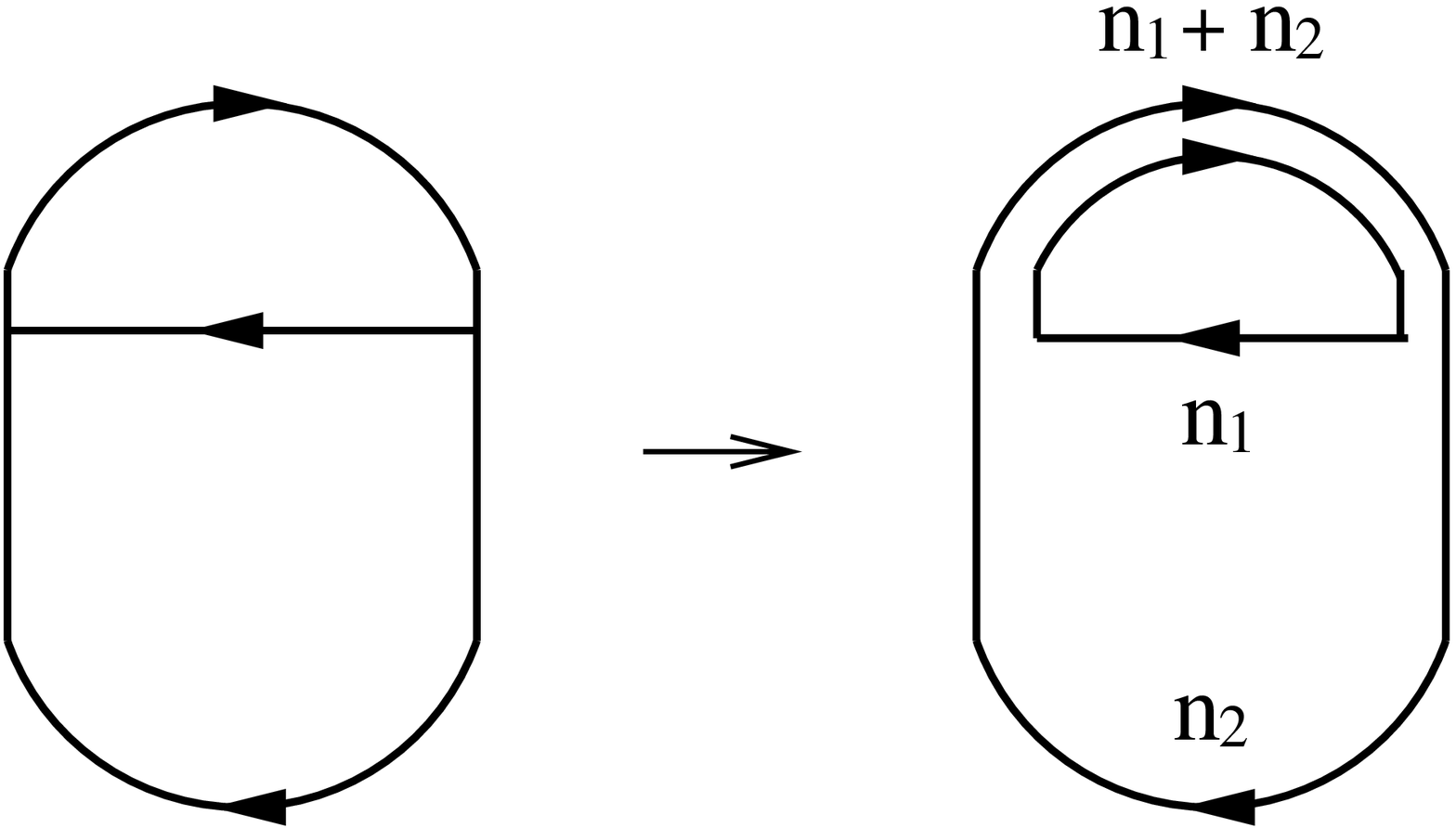}\\
\end{center}
\caption{}
\label{circuit2}
\end{figure}

\begin{figure}
\begin{center}
 \leavevmode
 \epsfxsize=70mm
 \epsfbox{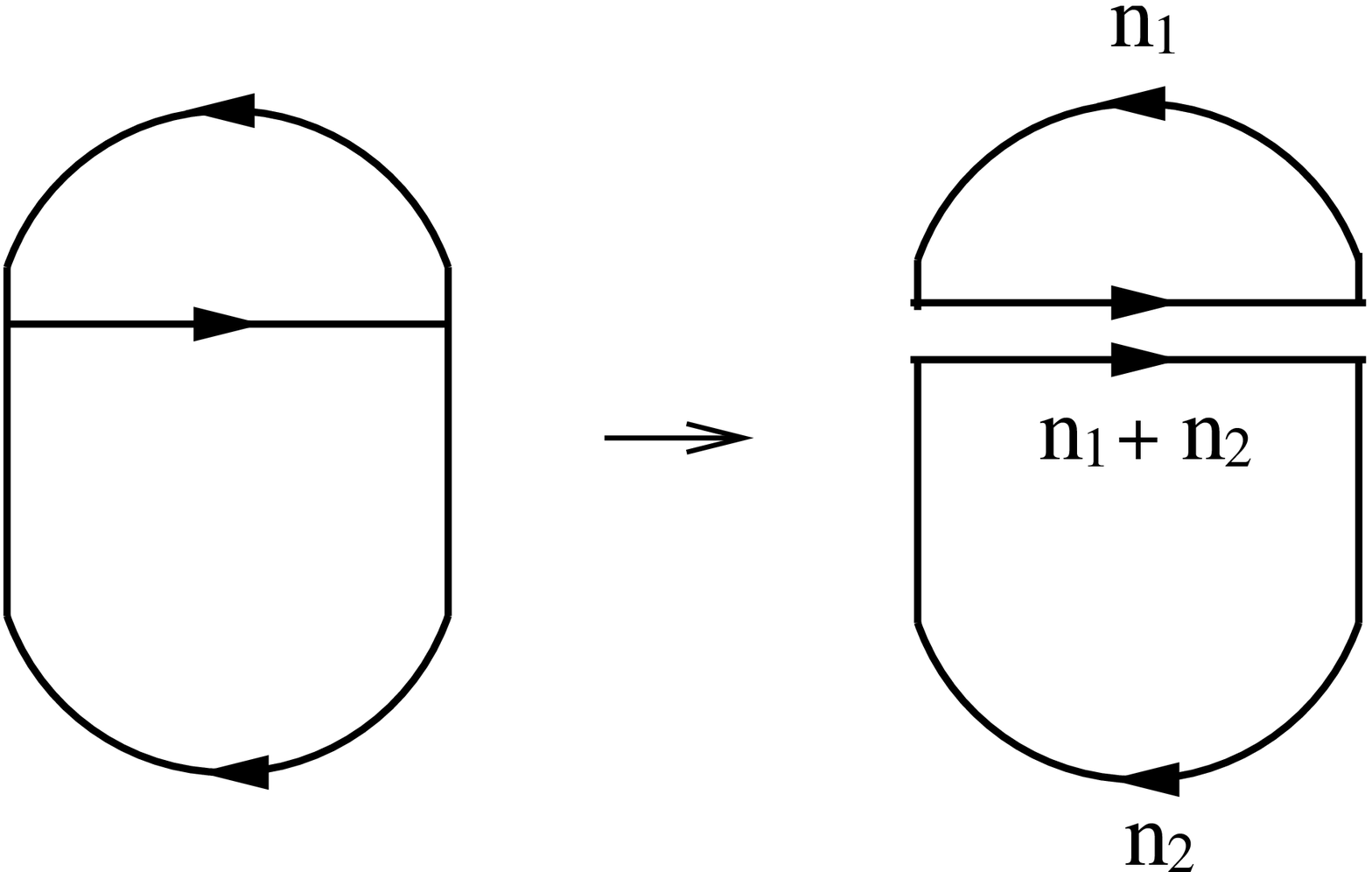}\\
\end{center}
\caption{}
\label{circuit3}
\end{figure}

\begin{figure}
\begin{center}
 \leavevmode
 \epsfxsize=40mm
 \epsfbox{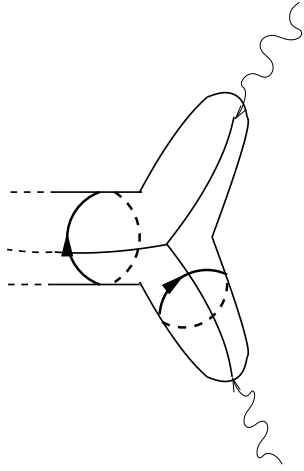}\\
\end{center}
\caption{An example of
a closed string tree diagram
in which the circuit of the type
Fig.\ref{circuit1} appears.
If one draws a corresponding gauge theory
Feynman diagram 
on a plane (view the above 
worldsheet from the right),
the cut-out loops look as depicted 
in the righthand side of Fig.\ref{circuit1}.
The closed circuit in the lefthand side of
Fig.\ref{circuit1}
can be interpreted as 
a limit where
two cut-out loops 
(partially) overlap.}
\label{t1}
\end{figure}

\begin{figure}
\begin{center}
 \leavevmode
 \epsfxsize=40mm
 \epsfbox{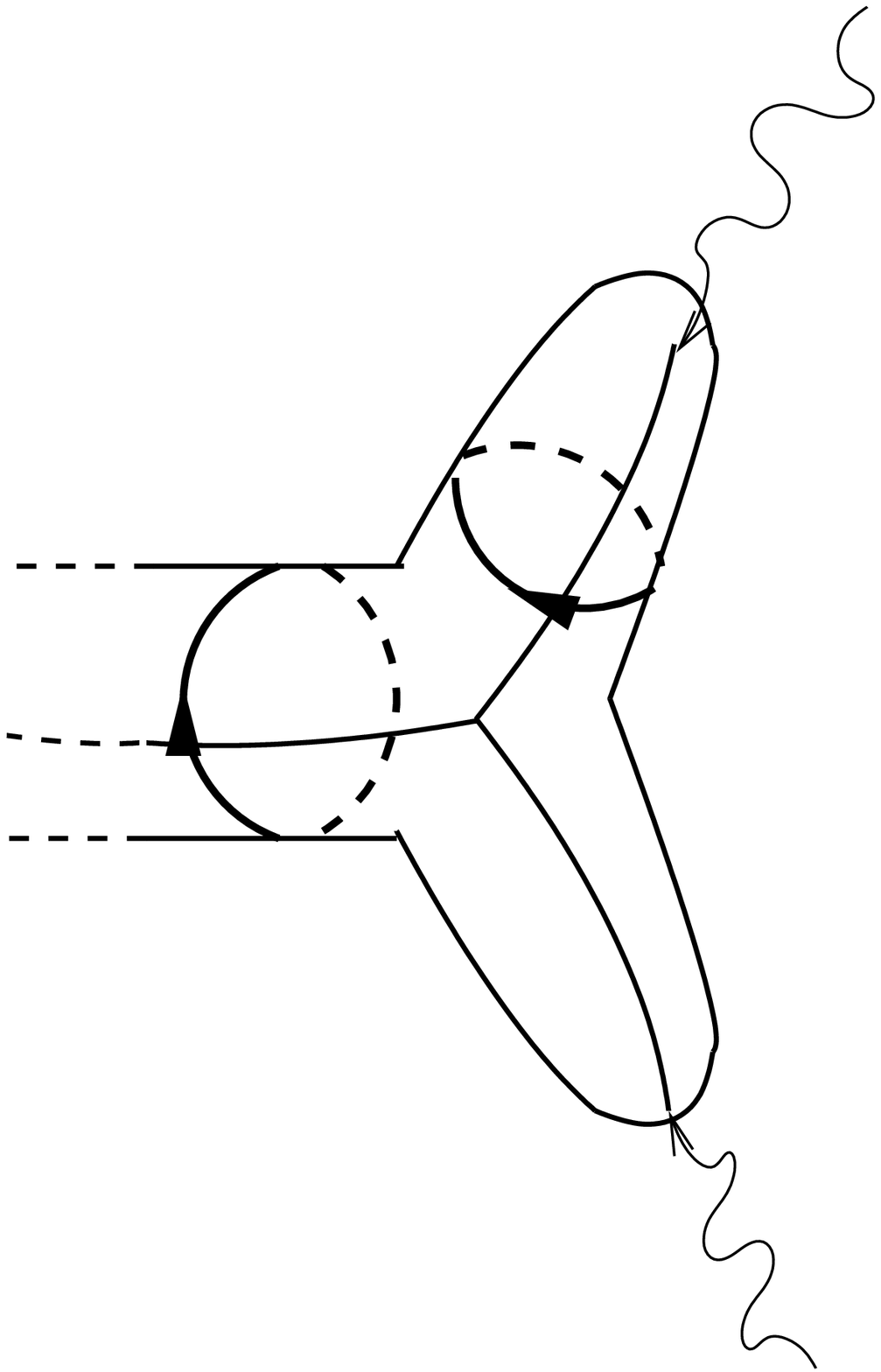}\\
\end{center}
\caption{An example of a 
closed string tree diagram
in which the circuit of the type
Fig.\ref{circuit2} appears.}
\label{t2}
\end{figure}

\begin{figure}
\begin{center}
 \leavevmode
 \epsfxsize=40mm
 \epsfbox{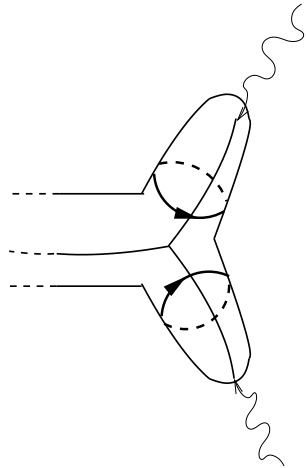}\\
\end{center}
\caption{An example of a 
closed string tree diagram
in which the circuit of the type
Fig.\ref{circuit3} appears.}
\label{t3}
\end{figure}

So far, I have studied the case
in which cut-out loops are isolated.
In general, 
bold arrows make up closed circuits:
More than two bold arrows can come
in and go out at one index-momentum loop 
(Fig.\ref{circuit})
to give a non-vanishing contribution. 
This case can be treated as an overlap
of single cut-out loops.
To see this, it
may be useful to draw an
analogy between 
the bold arrow circuits
with Feynman diagrams:\footnote{%
The analogy here is with general Feyman diagrams,
not with the specific large $N$ 
Feynman diagrams which have been studied
in this article.}
The bold arrows can be put on
the edges of the dual graph of
a 't Hooft-Feynman diagram.
The dual graph can be obtained
by replacing the faces of the 
original graph by dual vertices,
the edges by orthogonal dual edges and
the vertices by dual faces
(Fig.\ref{circuit}).
The bold arrows
make up closed circuits
on the dual graph of the 
planar Feynman diagram.
When one 
Fourier expands the correlation
function in terms of the gauge 
index dependent factor
$e^{-2\pi i \frac{a_i}{N}}$,
the bold arrow
tells whether
the phase $2\pi i \frac{a_i}{N}$ 
from the propagator
which it crosses
appears with positive sign
or negative sign.
In the limit $N \rightarrow \infty$,
$\frac{a_i}{N} \rightarrow \theta_i$,
which was implicit in the above,
one can regard the Fourier mode 
as an analogue of discrete ``momentum".
Then, each index sum 
picks up a term
in which all of the phases cancel.
This is an analogue of the
momentum conservation
at each vertex on Feynman diagrams.
One can solve the ``momentum conservations"
to end up with ``loop momenta" for
sub-loops of the circuit, see
Fig.\ref{circuit1}-\ref{circuit3}.
Note that I have specified the
direction of the bold arrows,
so the ``momentum"
should be always positive, i.e.
either
positive incoming 
or positive outgoing.
The circuits with
the same topology but with different
directions of arrows
should be regarded as different circuits.
Each sub-loop on the cut-out circuit
has a 
direction inherited from the
direction of the
bold arrows of the cut-out circuit, 
in particular
must cross an edge of the tree sub-diagram
which expresses the external momentum flow,
as in the case of the single 
cut-out loop.
Thus as before it can be associated with
an edge of a closed string tree diagram.
The energy dependent factor
of each sub-loop 
is the same
as that of a single cut-out loop.
Therefore, one can interpret such
a sub-loop as a single cut-out loop:
Closed circuits
are interpreted as made of
overlapping cut-out loops
(Fig.\ref{t1}-\ref{t3}).


As an example, below
I
study the case where
two cut-out loops
overlap at one index loop (Fig.\ref{circuit}),
again in the free limit.
The external momenta
entering into cut-out loops are
$k_{01}$ and $k_{02}$.
I assume that both of the
cut-out loops are 1-1 cut out loops,
other cases can be treated similarly.
These two cut-out loops give a contribution 
proportional to the factor
\bea
 \label{2cc11}
&&
\sum_{a_1=1}^N \cdots \sum_{a_J=1}^N
\sum_{b_1=1}^N \cdots \sum_{b_J=1}^N
\prod_{i=1}^{J_1}
\prod_{j=1}^{J_2}
\delta_{a_1b_1}
\frac{1}%
{e^{\beta \w + 2\pi i \frac{a_i-a_{i+1}}{N}} -1}
\cdot
\frac{1}%
{e^{\beta \w + 2\pi i \frac{b_j-b_{j+1}}{N}} -1}\nn
&=&
\sum_{a_1=1}^N \cdots \sum_{a_J=1}^N
\sum_{b_1=1}^N \cdots \sum_{b_J=1}^N
\prod_{i=1}^{J_1}
\prod_{j=1}^{J_2}
\delta_{a_1b_1}
\sum_{n_i=1}^\infty
\sum_{m_j=1}^\infty
e^{-n_i(\beta \w + 2\pi i \frac{a_i-a_{i+1}}{N})}
e^{-m_j(\beta \w + 2\pi i \frac{b_j-b_{j+1}}{N})}\nn
&=&
\sum_{n=1}^{\infty} e^{- n\beta \w J_1} \cdot
\sum_{m=1}^{\infty} e^{- m\beta \w J_2}
\nn
&=&
\frac{1}{e^{\beta \w J_1} -1}
\cdot
\frac{1}{e^{\beta \w J_2} -1}  .
\eea
Here, $J_1$ and $J_2$ 
are the numbers of the bold arrows
on the cut-out loops and
$a_1=b_1$ is the index of
the index loop
where two cut-out loops overlap.
From the second line to the third line
of (\ref{2cc11}), the 
``momentum conservation" 
was explicitely solved
by using the ``loop momenta"
$n$ and $m$.
Eq.(\ref{2cc11}) is the correct factor
for the contributions from two cuts
in the real time formulation of closed string
field theory at finite temperature.
One can also obtain the on-shell
delta functions 
$\delta(k_{01}^2-(J_1\w)^2)$ and
$\delta(k_{02}^2-(J_2\w)^2)$,
as in the case of a single cut-out loop.

\subsection{Closed string interaction vertices}

\begin{figure}
\begin{center}
 \leavevmode
 \epsfxsize=110mm
 \epsfbox{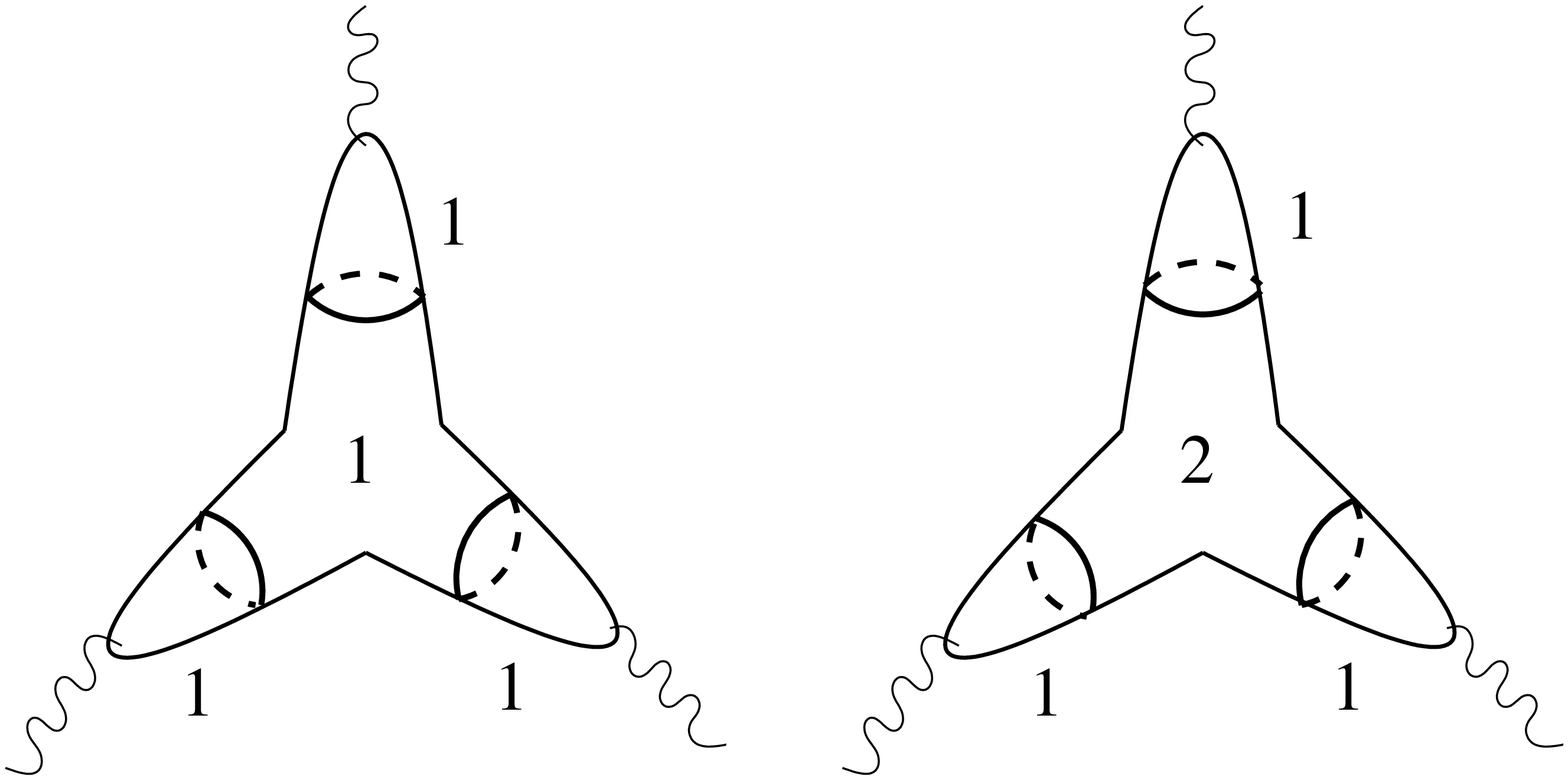}\\
\end{center}
\caption{Feynman diagrams divided
into type-1 regions and type-2 regions
by the cut-out loops.
The type-2 region in the diagram on the right
is complex conjugate to the corresponding
type-1 region in the diagram on the left.}
\label{closedv}
\end{figure}

\begin{figure}
\begin{center}
 \leavevmode
 \epsfxsize=100mm
 \epsfbox{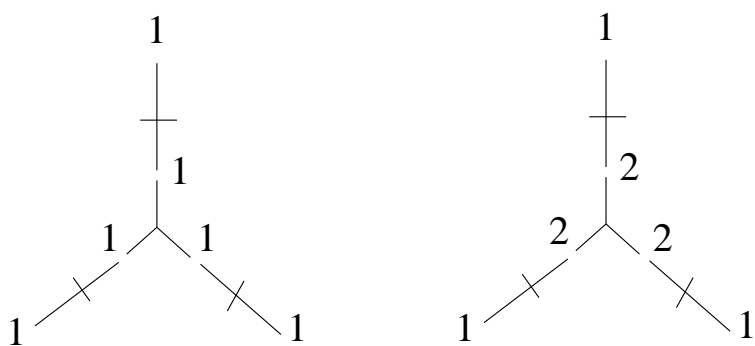}\\
\end{center}
\caption{The diagrams in Fig.\ref{closedv} drawn
so that they look as diagrams in the
real time formulation of
finite temperature field theories.}
\label{clvreal}
\end{figure}

Next, I argue that 
the interaction vertices
for the closed string field theory
arising from the gauge theory 
Feynman diagrams
also have the desired property
as required by the real time formalism.
Recall that each cut out
region  are either 
made of entirely type-1 or entirely type-2
propagators and vertices.
Also recall that the 2-2 propagator is the complex conjugate
of the 1-1 propagator (see (\ref{D0})):
\bea
i D^{(22)}_0 (k_0) = (i D^{(11)}_0(k_0))^{*} .
\eea
Similarly, for the 
type-1 and 
type-2 interaction vertices (\ref{rp}):
\bea
i\, \tr V_2[\Phi_{(2)}] 
= 
\left.(i\, \tr V_1[\Phi_{(1)}])^{*}
\right|_{\Phi_{(1)}\rightarrow \Phi_{(2)}}.
\eea
For a given diagram with
a cut-out type-2 region,
there is a corresponding diagram
which is obtained by
replacing all
the type-2 propagators 
and vertices in the region
by those of type-1 (Fig.\ref{closedv},\ref{clvreal}).
Then, the contribution
of the type-2 region
is the complex conjugate
of the corresponding type-1 region.
This means that
the type-2 vertices
of the closed string field theory 
obtained from the gauge theory 
are the 
complex conjugate of 
the corresponding type-1 vertices.
This is 
the property of the interaction vertices
in the real time formulation of 
finite temperature field theories.

Finally, recall that in each cut-out region
there are no further cuts, i.e. no
further temperature dependent piece.
This means that these cut-out regions
probe the zero-temperature geometry
of the bulk, i.e. AdS$_5$ for
the four dimensional ${\cal N}=4$ super Yang-Mills
theory on $S^3$.\footnote{%
Since I have been studying a quantum mechanics
dimensionally reduced from four dimensions,
strictly speaking I have only considered
the s-wave for the $S^3$ part of 
the coordinates on AdS$_5$.
But generalization to include excited modes
on the $S^3$ will be straightforward.}

Thus altogether, the real time formulation
of the gauge theory at finite temperature
in the confined phase describes the
real time formulation of 
classical closed string field theory on AdS
at finite temperature.

\section{Summary and Discussions}\label{Summ}

\subsection*{\it Summary}

Successful tests of the AdS-CFT conjecture
encourage our hope that
it has the answers to all the puzzles
surrounding black holes,
at least in asymptotically AdS spaces.
However, to extract such answers,
one needs to understand 
how to translate closed string descriptions
to gauge theory descriptions
and vice versa, 
in particular
what distinguishes 
between the geometries with and without a black hole
in the gauge theory descriptions.
And to solve real puzzles about black holes,
one must work with the Lorentzian signature.

In this article,
I have given a prescription
for extracting the 
dual bulk description
corresponding to the confined phase
in the real time formulation of
gauge theories at finite temperature.
With this prescription,
I have shown 
how the difference between the phases,
namely the confined and the deconfined phases,
changes the relevant Feynman diagrams
in the planar limit.
In the confined phase,
the Feynman diagrams of the gauge theory 
organize themselves into tree diagrams of
closed string field theory 
on AdS at finite temperature
in the real time formalism.

\subsection*{\it Study from the closed 
string side}

In this article, I have 
compared the gauge theory results 
with the quite 
general proparties of
the real time formulation of the 
closed string field theory.
It is certainly interesting to re-examine
the real time formulation
of closed string field theory
at finite temperature
\cite{Leblanc:1987zj} in more detail,
in comparison with the analysis of the
large $N$ Feynman diagrams in this article.%
\footnote{In the BMN
plane wave limit \cite{Berenstein:2002jq},
the real time formulation of light-cone 
string field theory
was recently studied in \cite{Abdalla:2005qs}.
However, note that the BMN limit is a 
different limit from the 't Hooft limit.}
Since what one directly obtains from
the 't Hooft-Feynman diagrams are closed string
worldsheets rather than the
string field theory, it may also be useful
to investigate the worldsheet formulation
of the real time formalism \cite{Mathur:1993tp}.
In particular, it will be nice
if one can characterize the cut-out loop
in terms of the closed string worldsheet language,
and see the parallel between the results in this article.%
\footnote{In the case of a worldsheet with a torus topology, 
the cut-out loop (in my terminology)
was characterized by a boundary condition 
\cite{Abdalla:2004dg}.}

\subsection*{\it Deconfinement, Hawking-Page and the
Kosterlitz-Thouless transition}

In the deconfined phase,
a different type of Feynman diagram,
namely a diagram with
unrestricted cuts, 
becomes relevant.
This difference
should be reflecting the distinction between
the geometry with and without a black hole,
as long as the identification
of confinement-deconfinement phase transition
with the Hawking-Page transition
is correct.
This may be regarded as a Lorentzian
signature counterpart of the
Kosterlitz-Thouless phase transition
on the string worldsheet
\cite{Sathiapalan:1986db,Kogan:1987jd,Atick:1988si}.
There the condensation of the winding modes
makes the string worldsheet interpretation
difficult.
The mechanism
that is supposed to be
corresponding to 
the Kosterlitz-Thouless phase transition
in the large $N$ Feynman diagrams
was discussed in \cite{Furuuchi:2005qm} 
(see also
\cite{Kazakov:2000pm}).
It has long been anticipated 
that the Kosterlitz-Thouless phase transition
on the string worldsheet
is analogous to the deconfinement phase
transition where underlying
degrees of freedom manifests itself.
It is intriguing that 
in the AdS-CFT correspondence this is
not just an analogue 
but exactly the same thing (the equivalent dual description). 
The endpoint of the phase transition
may still admit a worldsheet interpretation
in a {\em different} geometry
\cite{Furuuchi:2005qm,KalyanaRama:1998cb,%
Barbon:2001di,Barbon:2002nw,%
McGreevy:2005ci},
however, this point needs further study.\footnote{%
Historically, string theory started as a description
of confinment. 
From this point of view, the correspondence
between the deconfined phase and the black hole geometry
makes it highly non-trivial if
the black hole geometry still allows a string description.
It might be the case that such a description is possible
only for a specific class of observers (coordinate frames),
as suggested by the black hole complementarity
\cite{Susskind:1993if}.}
Note that the Euclidean case
studied in the references mentioned above
is somewhat
simplified situation compared with
the original Lorentzian problem,
since the former only contains
the region outside the black hole horizon.

\subsection*{\it The role of the imaginary time formalism}

The prescription
I have given for
reading off the dual description of 
the confined phase
uses  
the Matsubara contour
to determine the boundary conditions for
the thermal Green's function.
In this sense,
the real time formalism is not completely
different from the imaginary time
formalism but rather it contains
the latter in the imaginary direction of the contour.
The relevance of the Matsubara contour
for determining the boundary condition
reminds us of the role of the
Euclidean path integral in
the models of gravitational physics
\cite{Hartle:1976tp,Gibbons:1976ue,Hartle:1983ai}
(see also
\cite{Maldacena:2001kr}).

In the gauge theory context,
the relevance of the Matsubara contour
indicates that
the methods which are useful in
the imaginary time formalism
may have direct relevance
also in the real time formalism.
In particular,
the crucial role played by
the temporal component of the
gauge field in this article seems
to suggest that 
the 
Polyakov's criterion for confinement
\cite{Polyakov:1978vu}
has some simple extension
to the real time formalism.
Note that 
the boundary
condition is not just a simple consequence
of the projection onto the gauge singlet sector,
but also includes the saddle point calculation.
It is important to further investigate
this role of the imaginary time
formalism in the real time formalism.

\subsection*{\it On dynamical formation of black holes}

How a black hole is formed 
from ordinary matter
is an important question,
which is
also relevant for
the information loss paradox \cite{Hawking:1976ra}.
I would like to point out that 
to describe such a process,
the
initial geometry 
without the black hole
should be probed
by the prescription I have given,
namely by taking into account the 
configuration of the temporal component $A_0$
of the gauge field,
characteristic of the confined phase,
on the vertical parts of the contour.
Moreover the final state,
the large black hole in AdS space,\footnote{%
Large here means the larger of
the two Schwarzshild-AdS
solutions in \cite{Hawking:1982dh}.}
should correspond to
the gauge theory in
the deconfined phase.

Formation of a black hole
is a dynamical process,
and it may be worth mentioning 
that
the real time formalism
has a close relation with the
Schwinger-Keldish technique
for describing non-equilibrium systems
\cite{Schwinger:1960qe,Keldysh:1964ud}.
The techniques
developed in this article
may also find application in the study of
non-equilibrium processes in
large $N$ gauge theories.
It will be interesting to apply the
Schwinger-Keldish technique
to the 
large $N$ gauge theories
to study the
dynamical formation of a black hole
via the AdS-CFT correspondence.

\subsection*{\it Statistical average and horizon}

In this article I have argued that
the real time formulation of
gauge theories at finite temperature
can describe the dual closed string field
theory on
AdS at finite temperature.
Together with 
the results in
the black hole phase \cite{Maldacena:2001kr},
the success in describing
{\em both} the black hole 
and the non-black hole phase
by the real time formulation
of gauge theory at finite temperature 
is an encouraging evidence for the 
AdS-CFT correspondence
in Lorentzian signature.
On the other hand, 
the real time formulation introduces
the doubled degrees of freedom
to describe the thermal ensemble
by entanglement.\footnote{%
This statement may be more 
appropriate for
the thermo field dynamics.
The Feynman rules derived 
from the thermo field dynamics 
are the same 
\cite{Semenoff:1982ev}
as those of
the real time formulation of
Ref.\cite{Niemi:1983nf}.}
The reason one usually
uses a statistical
average is that
one does not need to know
the state one is precisely in,
to describe the thermodynamical 
property of the system.
However, one expects that
there is actually a single pure state
at a given moment,
and
unitarity
may be understood more straightforwardly
without taking the statistical ensemble,
even though the real time formulation
of field theory at finite temperature is
unitary.
It should be interesting to 
understand more
precisely what actually the
statistical average in the gauge theory side
introduces in the bulk geometry.
This question may be related to 
the recent proposal of \cite{Mathur:2003hj}.
See also \cite{Maldacena:2001kr}.

\subsection*{\it $1/N$ effects}

In this article I worked in the
leading order in the $1/N$ expansion.
This corresponds to the classical
closed string theory in the bulk.
This was sufficient
since the argument from the
Carter-Penrose diagram is 
based on
classical gravity.
However, it is also important
to study
how the $1/N$ corrections, which
corresponds to quantum 
corrections in the bulk, 
modify the above classical view.%
\footnote{The role of the $1/N$ corrections 
in a related context was
stressed in \cite{Festuccia:2005pi}.
The sum over contributions
from both the AdS-Schwarzshild black hole
geometry and the AdS geometry
discussed in \cite{Maldacena:2001kr}
is a non-perturbative effect in $1/N$.
Note that even though the contributions 
must be summed over,
each saddle point in the gauge theory
should have a dual bulk interpretation,
at least at large $N$.
The objective of this article has been to
read off, or ultimately
{\em derive}, the bulk geometry
from the gauge theory.}
Also,
the information paradox arose from the
discovery of
Hawking radiation \cite{Hawking:1974sw}
which is a quantum effect.
The method developed in \cite{Brigante:2005bq}
may be useful for the investigation in
this direction.

\subsection*{\it More puzzles 
than have been solved?}

In the confined phase, the
two-point Green's function between
a type-1 and a type-2 operator
should be identified 
with 1-2 propagator in closed string field theory.
This is quite natural
as a dual description of the confined phase
still at finite temperature.
In contrast,
despite some
successes in the comparisons 
between gauge theories in deconfined phase 
and black hole geometry in literature,
it is still not clear precisely
{\em how} correlation functions in
the large $N$ gauge theory 
describe the dual theory on
the black hole geometry.
This raises a question 
as to how 
the dual of deconfined phase
is encoded in the gauge theory
correlation functions.
In the spirit followed
in this article for the confined phase,
one should also be able to describe
from the gauge theory
how
the type-2 fields are
organized in the deconfined phase
into the
degrees of freedom behind the
horizon.
Thus the answer to the question
raised in the introduction:
``How does confinement change the role of
type-2 fields in the bulk?"
now comes back as a question to
the original interpretation:
{\em Precisely how does deconfinement
change the role of
the type-2 fields in the bulk?} \footnote{%
I thank J. R. David for raising this question.}

\vspace*{6mm}
\begin{center}
{\bf Acknowledgments}
\end{center}
\vspace*{-1mm}
It is my pleasure to 
express my thanks to 
D. Astefanesei,
J. R. David,
D. Ghoshal,
R. Gopakumar,
D. P. Jatkar,
S. Naik,
A. Sen,
K. Sengupta
and K. P. Yogendran
for useful discussions and valuable comments
which substantially improved
the quality of this article.
I am also grateful to D. Ghoshal, N. Mahajan and 
K. P. Yogendran for careful
reading of the manuscript and 
various useful suggestions. 
I sincerely appreciate
liberal support for our research
from the people in India.

\appendix

\section*{Appendix}

\section{A sample calculation}\label{A}

\begin{figure}
\begin{center}
 \leavevmode
 \epsfxsize=90mm
 \epsfbox{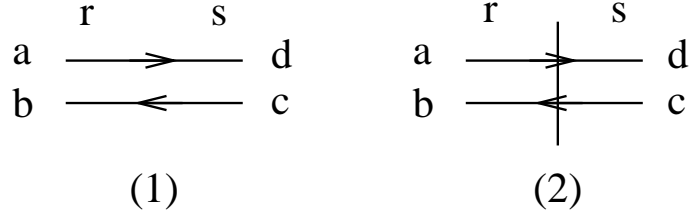}\\
\end{center}
\caption{Feynman propagators. $a,c$ 
refer to the
first gauge index of the adjoint field
$\Phi_{(r)ab}$, and 
$b,d$ refer to the second.
$r,s=1,2$ refer to type-1 and type-2.}
\label{AppFeyn}
\end{figure}

\begin{figure}
\begin{center}
 \leavevmode
 \epsfxsize=80mm
 \epsfbox{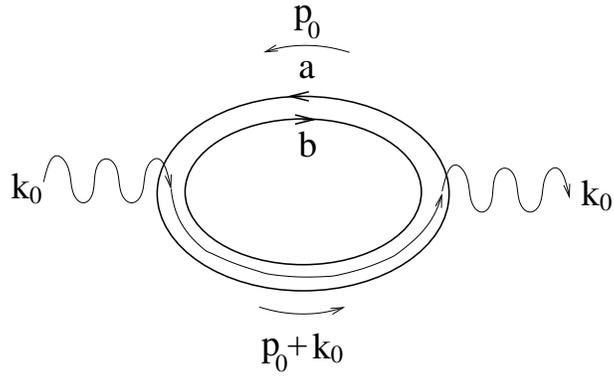}\\
\end{center}
\caption{A Feynman diagram with no cut contributing to
$\langle \tr \Phi^2_{(1)}(k_0) 
\tr \Phi^2_{(1)}(-k_0) \rangle$.}
\label{samp}
\end{figure}

\begin{figure}
\begin{center}
 \leavevmode
 \epsfxsize=80mm
 \epsfbox{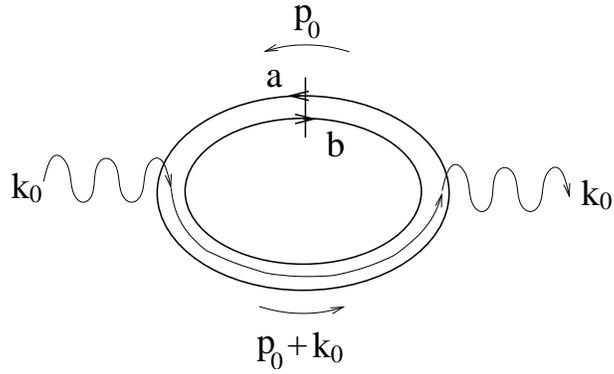}\\
\end{center}
\caption{The one-cut contribution to 
$\langle \tr \Phi^2_{(1)}(k_0) 
\tr \Phi^2_{(1)}(-k_0) \rangle$.}
\label{samp1}
\end{figure}

\begin{figure}
\begin{center}
 \leavevmode
 \epsfxsize=80mm
 \epsfbox{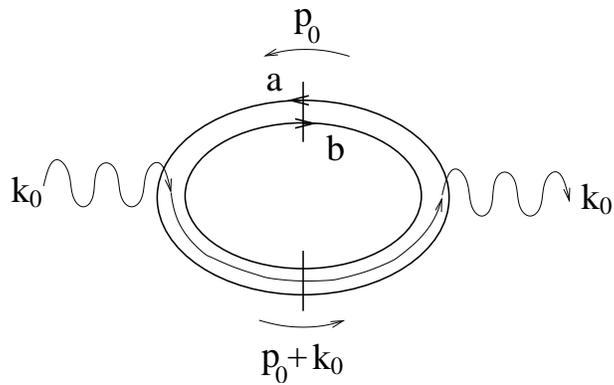}\\
\end{center}
\caption{The
two-cut (cut-out) contribution to 
$\langle \tr \Phi^2_{(1)}(k_0) 
\tr \Phi^2_{(1)}(-k_0) \rangle$.}
\label{samp2}
\end{figure}

In this appendix, I provide a simple example 
to 
illustrate how the mechanism
which selects the surviving
Feynman diagrams in the confined phase works.
I calculate
the two-point function
$\langle \tr \Phi^2_{(1)}(k_0) 
\tr \Phi^2_{(1)}(-k_0) \rangle$
(the total momentum conservation
delta function has been dropped),
where $\Phi_{(1)}$ is the type-1
adjoint scalar field,
in the free limit of the $SU(N)$ 
gauge 
quantum mechanics.
The case for the
correlation functions
with type-2 fields are similar.
The calculation follows the
explanation in the main text,
and should be compared with the 
general arguments there.
This simple example still
captures the essential points
of the mechanism which selects the
non-vanishing diagrams in the confined phase.
In the following,
all the 
overall constants are omitted since they are
not important for the present purpose. 
Since I am studying the free theory
in this example,
I only need to recall
the propagators (\ref{D0}) and (\ref{Db}).
These are depicted in Fig.\ref{AppFeyn}.
The first  
corresponds to the temperature independent part
\bea
i D_{0 ab,cd} =
\delta_{ad}\delta_{bc}
\left(
\begin{array}{cc}
 \frac{i}{k_0^2-\w^2+i\e} & 0 \\
 0 & \frac{-i}{k_0^2-\w^2-i\e}
\end{array}
\right).
\eea
The matrix structure 
of the propagator
originates
from the fact that one needs to
include two types of fields,
type-1 and type-2, for describing
the real time formulation of 
the finite temperature
field theory.
The (1-1) matrix component corresponds to
the type-1-type-1 propagator, and so on.
The second in Fig.\ref{AppFeyn}
corresponds to the temperature dependent part
of the propagator
\bea
i D_{\beta ab,cd} &=&
\delta_{ad}\delta_{bc}
\pi \delta(k_{0}^2- \w^2) \nn
&&
\frac{1}{e^{{|\beta k_0+2\pi i\frac{a-b}{N}}|_R}-1}
\left(
\begin{array}{cc}
 1 & e^{\frac{1}{2}|\beta k_{0}+2\pi i\frac{a-b}{N}|_R} \\
e^{\frac{1}{2}|\beta k_{0}+2\pi i\frac{a-b}{N}|_R} & 1
\end{array}
\right) .
\eea
The temperature
dependent part of the propagator
is drawn with the ``cut",
i.e. the vertical line in the 
Fig.\ref{AppFeyn} (2).
This cut is one of the main tools
in the following discussions.
The relevant Feynman diagrams are
depicted in Fig.\ref{samp}-\ref{samp2}.
The total momentum flowing
on a propagator
is a sum of momenta on
its two index lines
(taking into account the
signs indicated by the arrows), 
and
external momentum 
($k_0$ in this case)
flowing
between the double lines 
if there is any.
Fig.\ref{samp} 
is the temperature
independent contribution
and probes the zero-temperature
bulk geometry.
The calculation is standard so
I omit the explanation of this case,
see e.g. 
\cite{Gopakumar:2003ns,Gopakumar:2004qb,%
Gopakumar:2005fx,Gopakumar:2004ys,%
Furuuchi:2005qm}.\footnote{%
And see \cite{Son:2002sd}
for a prescription for the Lorentzian signature.}

\subsection{Only one cut on an index loop}

Fig.\ref{samp1}
has one cut, that is, the
temperature dependent part.
In this case, 
one can parameterize the loop integration variable
$p_0$ so that the external momentum
flows through the propagator
without the cut, as shown in  Fig.\ref{samp1}.
This Feynman diagram gives
\bea
\sum_{a=1}^N
\sum_{b=1}^N
\int dp_0
\frac{i}{p_0^2-\omega^2+i\e}
\delta((p_0+k_0)^2-\w^2)
\frac{1}{e^{|\beta(p_0+k_0)+2\pi i\frac{a-b}{N}|_R}-1}.
\eea
In a planar diagram,
one can associate each loop momentum
to an index loop:
The number of the index loop is one more
than the number of the loop momentum,
but one index sum can be factored out.
Here, the sum over the index $b$ can be factored out 
by the shift $a \rightarrow a+b$ (mod $N$).
The rest of the indices are ``associated"
to the loop momentum,
in this case $a$ to $p_0$:
The gauge index and 
the loop momentum appear
in a specific combination
$\beta p_0 + 2\pi i \frac{a}{N}$.
After performing the $p_0$ integration
using the delta function,
one obtains (the overall factor $N$ 
coming from the sum over the index 
$b$ has been dropped)
\bea
&&
\frac{1}{2\w}
\sum_{a=1}^N
\frac{i}{(\w-k_0)^2-\omega^2+i\e}
\frac{1}{e^{|\beta\w+2\pi i\frac{a}{N}|_R}-1} \nn
&+&
\frac{1}{2\w}
\sum_{a=1}^N
\frac{i}{(\w+k_0)^2-\omega^2+i\e}
\frac{1}{e^{|-\beta\w+2\pi i\frac{a}{N}|_R}-1}.
\label{1cutdiagram}
\eea
The first term in 
(\ref{1cutdiagram})
contains a factor
\bea
 \label{avanish}
\sum_{a=1}^N
\frac{1}{e^{|\beta\w+2\pi i\frac{a}{N}|_R}-1}
=
\sum_{a=1}^N
\sum_{n=1}^\infty
e^{-n|\beta\w+2\pi i\frac{a}{N}|_R}
=0
\eea
and vanishes.
Recall that 
$|\cdots |_R$ was defined in (\ref{||R}) as
\bea
 \label{a||R}
|z|_R =
\left\{
\begin{array}{c}
z \quad (\mbox{Re}\, z > 0)\\
-z \quad (\mbox{Re}\, z < 0)
\end{array}
\right.   , \quad (\mbox{Re}\, z \ne 0).
\eea
More precisely,
I took $N$ to be strictly infinite
so that
the sum $\sum_{a=1}^N$
can be replaced with
the integral $N \int_0^1 d\theta$.
This picks out the constant mode
in the Fourier expansion
on the left hand side of (\ref{avanish}),
which is zero.
As explained in the main text,
(\ref{avanish}) is the basic
equation
which is relevant for the vanishing of 
a large class of Feynman diagrams
in the confined phase. 
The second term 
in (\ref{1cutdiagram})
also vanishes in 
the same way.


\subsection{The case in which 
a cut-out loop divides
the diagram into two pieces 
both containing an external leg}

\begin{figure}
\begin{center}
 \leavevmode
 \epsfxsize=65mm
 \epsfbox{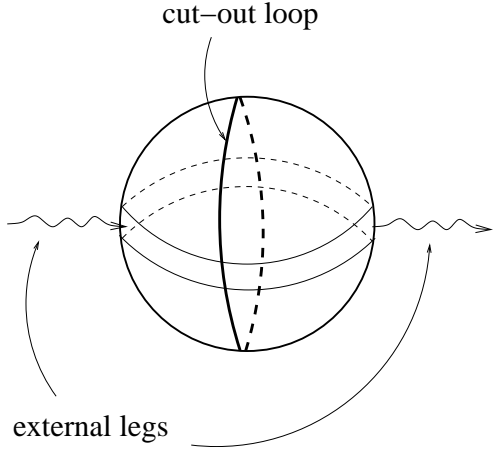}\\
\end{center}
\caption{The Feynman diagram 
in Fig.\ref{samp2}
drawn on a sphere.
The end points of the cuts are connected
if they are inside the same index loop, 
to make up a cut-out loop.
The cut-out loop divide the sphere into two regions
both containing an external leg.}
\label{sphere}
\end{figure}

Now let us turn to the
calculation of Fig.\ref{samp2}.
In Fig.\ref{samp2},
the two cuts 
divide the diagrams to two disconnected
pieces both containing an external leg.
If one connects the end points of the cuts
inside the same index loop, the cuts
make up a cut-out loop (Fig.\ref{sphere}).
The point is that in this case, 
the external momentum
flow 
must cross the cut, in contrast to
the previous case.
Since two regions 
of the diagram
are disconnected by
the cuts, one cannot avoid the
cut by a shift of loop integration variable $p_0$.
The Feynman diagram is calculated to be
\bea
\sum_{a=1}^N
\sum_{b=1}^N
\int dp_0
\delta(p_0^2-\w^2)
\frac{1}{e^{|\beta p_0+2\pi i\frac{a-b}{N}|_R}-1}
\delta((p_0+k_0)^2-\w^2)
\frac{1}{e^{|\beta(p_0+k_0)+2\pi i\frac{a-b}{N}|_R}-1}.
\eea
As in the previous case, one summation
over gauge indices factors out.
After performing the loop integral
one obtains
\bea
&&\sum_{a=1}^N
\frac{1}{e^{|\beta \w+2\pi i\frac{a}{N}|_R}-1}
\delta((\w+k_0)^2-\w^2)
\frac{1}{e^{|\beta(\w+k_0)+2\pi i\frac{a}{N}|_R}-1}
\label{p=w}
\\
&+&
\sum_{a=1}^N
\frac{1}{e^{|-\beta \w+2\pi i\frac{a}{N}|_R}-1}
\delta((-\w+k_0)^2-\w^2)
\frac{1}{e^{|\beta(-\w+k_0)+2\pi i\frac{a}{N}|_R}-1}
\label{p=-w}.
\eea

\subsubsection{The vanishing case}
\begin{figure}
\begin{center}
 \leavevmode
 \epsfxsize=80mm
 \epsfbox{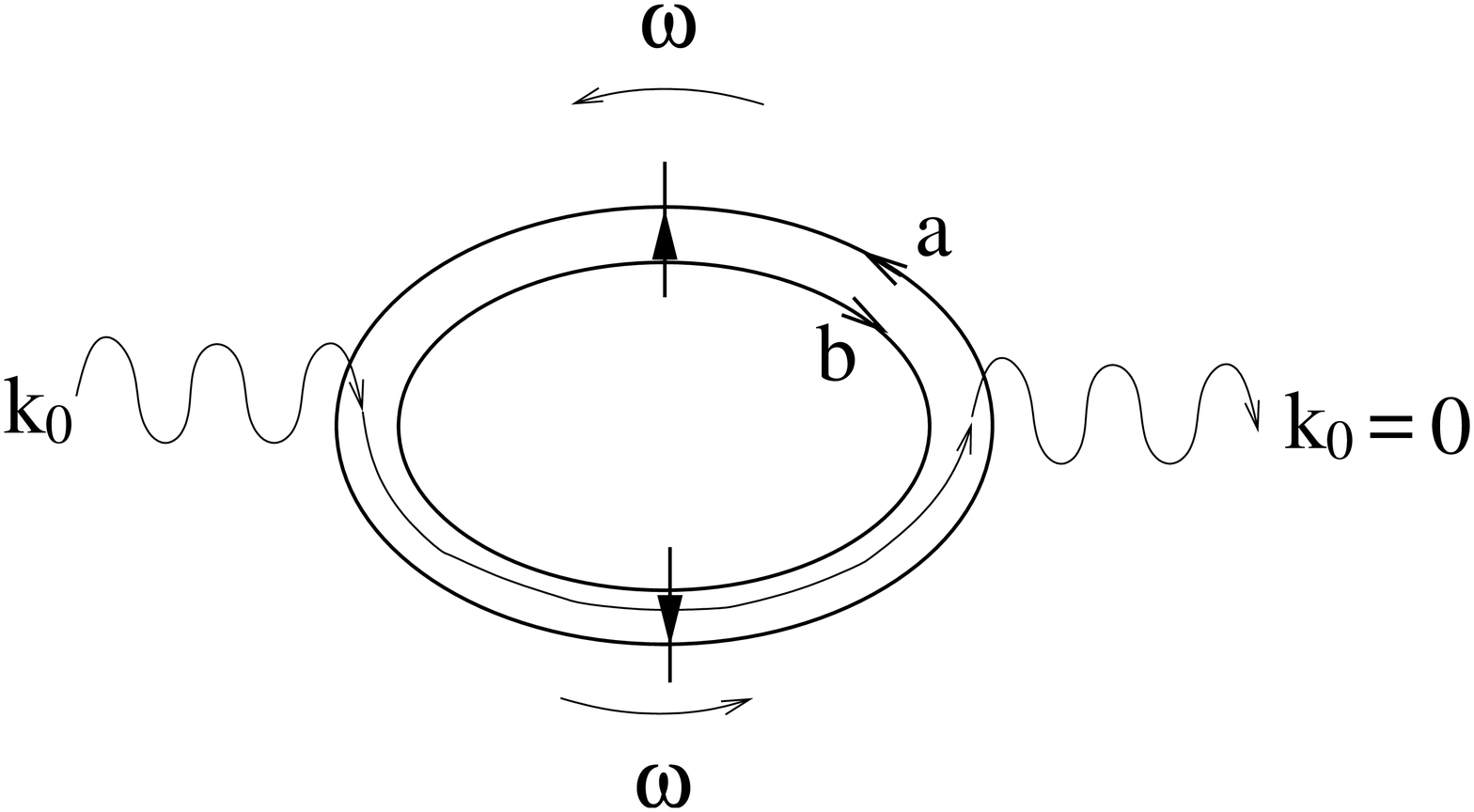}\\
\end{center}
\caption{The vanishing cut-out diagram.}
\label{sampv}
\end{figure}

The first line (\ref{p=w}),
a contribution from $p_0 = \w$ case,
can be divided into two cases:
$k_0 =0$ and $k_0 = -2\w$.
Fig.\ref{sampv} corresponds to the
$k_0 =0$ case.
\begin{figure}
\begin{center}
 \leavevmode
 \epsfxsize=75mm
 \epsfbox{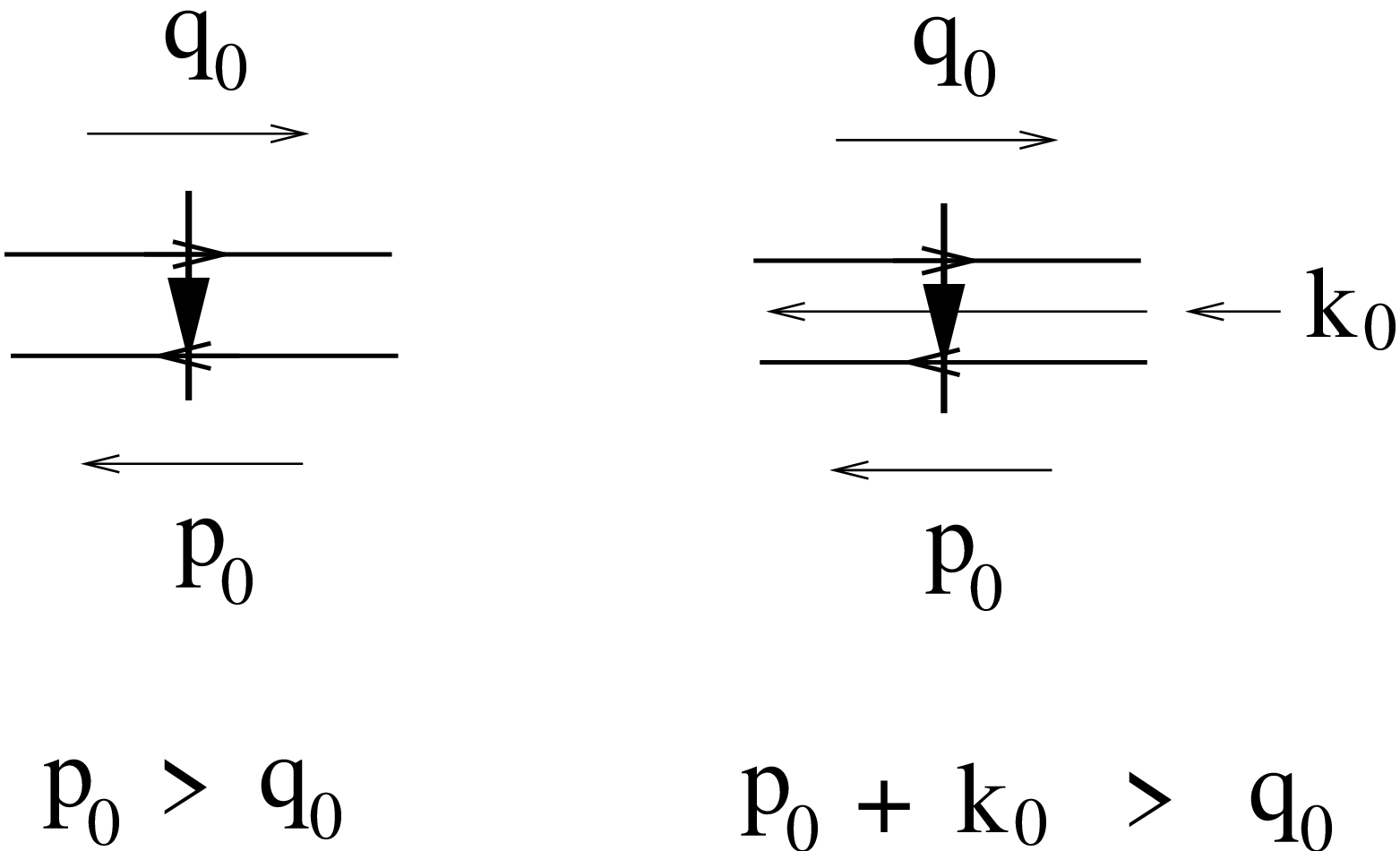}\\
\end{center}
\caption{A bold arrow 
put on a cut
indicates
which of the momenta associated
with the index is larger.
When there is an external momentum
flowing into a propagator
($k_0$ in the right),
it is added to the momentum with the same direction.}
\label{appboldarrow}
\end{figure}
Following the rule given in
Fig.\ref{appboldarrow},
I put bold arrows on the cuts
to indicate which of the directed 
momenta
is larger. 
The external momentum flow
is added to the loop momentum
in the same direction.
In other words,
the direction of the
bold arrow is determined
according to the direction
of the total momentum
flowing on a propagator.
The momentum associated with
the factored out index
is regarded as zero.
In Fig.\ref{sampv},
the bold arrows are always directed from the 
$b$-index line to $a$-index line.
This means that the momenta on the
$a$-index line is always bigger
than that of the $b$-index line
(which is regarded as zero, as 
mentioned above).
Therefore, 
Fig.\ref{sampv} is a contribution to 
Fig.\ref{samp2}
in the following situation:
\bea
&&
\left|\beta p_0+2\pi i\frac{a}{N}\right|_R
= \beta p_0+2\pi i\frac{a}{N} \label{plus1}\\
&&
\left|\beta(p_0+k_0) +2\pi i\frac{a}{N}\right|_R
=\beta(p_0+k_0)+2\pi i\frac{a}{N}, \label{plus2}
\eea
where $|\cdots |_R$ was defined in (\ref{a||R}).
Thus I obtain
\bea
&&
\frac{1}{2\w}
\sum_{a=1}^N
\frac{1}{e^{|\beta \w+2\pi i\frac{a}{N}|_R}-1}
\delta(k_0)
\frac{1}{e^{|\beta \w+2\pi i\frac{a}{N}|_R}-1} \nn
&=&
\frac{1}{2\w}
\delta(k_0) 
\sum_{a=1}^N
\sum_{n_1=1}^\infty
e^{-n_1 (\beta \w+2\pi i\frac{a}{N})}
\sum_{n_2=1}^\infty
e^{-n_2 (\beta \w+2\pi i\frac{a}{N})}
=0.
\eea
The point is that
the sign of the phase $2\pi i\frac{a}{N}$
in the Fourier expansion is always  
the same.
This is a consequence of
(\ref{plus1}) and (\ref{plus2}),
which are indicated by the bold arrows
in Fig.\ref{sampv}.
There are no cancellations
of the $a$-dependent phases
and hence it vanishes
upon summation over gauge index $a$,
like in the previous case (\ref{avanish}).

\subsubsection{The non-vanishing case}
\begin{figure}
\begin{center}
 \leavevmode
 \epsfxsize=80mm
 \epsfbox{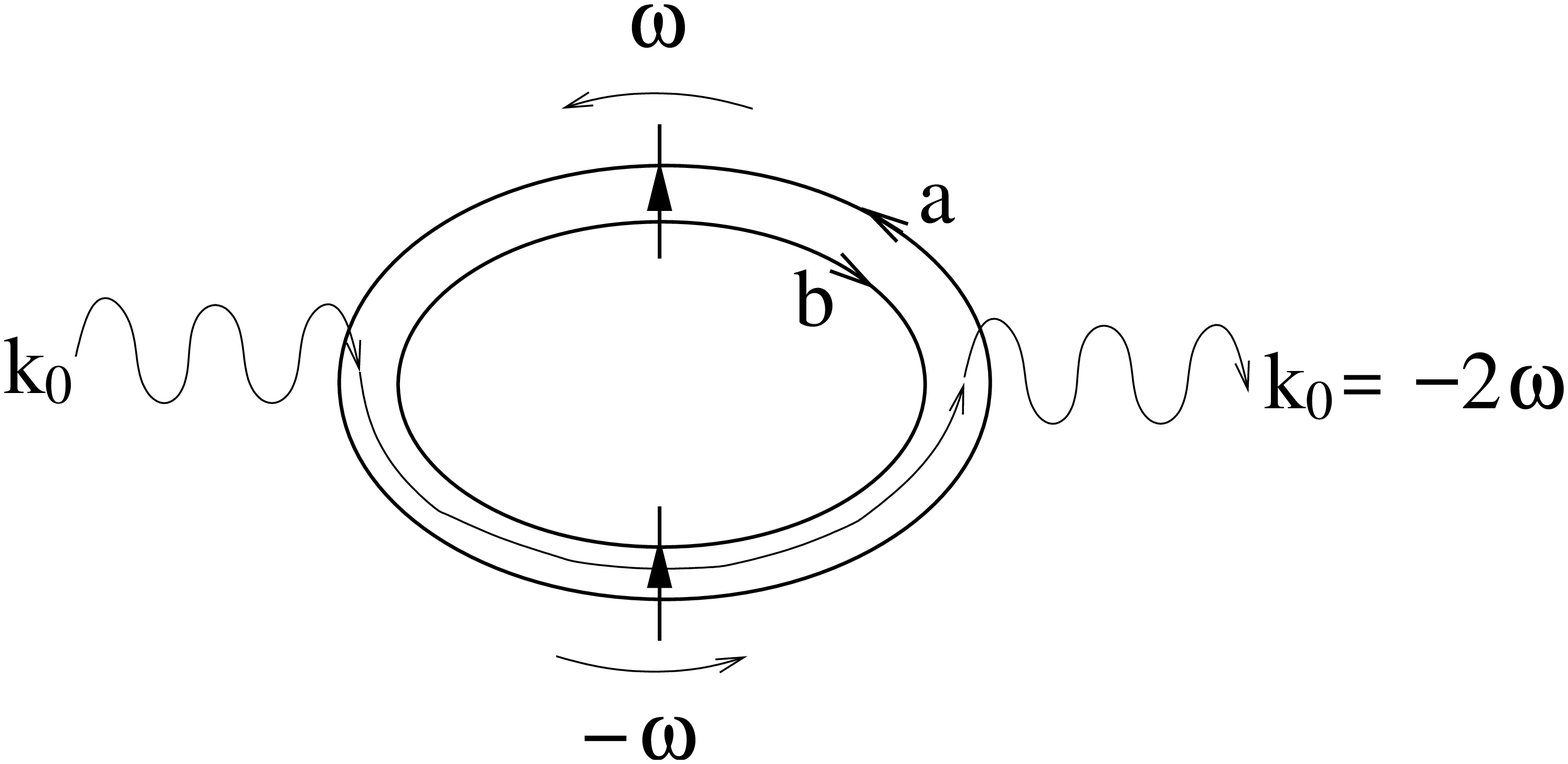}\\
\end{center}
\caption{The non-vanishing cut-out diagram.}
\label{sampnv}
\end{figure}

Now let us turn to the $k_0= -2\w$ case
in (\ref{p=w}).
This corresponds to Fig.\ref{sampnv}.
In this case,
in the upper propagator
the bold arrow is directed
from the $b$-index line to
$a$-index line,
whereas in the lower
propagator
the direction is reversed.
This means that the momentum in
the direction of
$a$-index line is bigger
than that in the direction of the $b$-index line
in the upper propagator,
and smaller
in the lower one (Fig.{\ref{appboldarrow}}).
Therefore, 
Fig.\ref{sampnv} is a contribution to
Fig.\ref{samp2} in the following situation:
\bea
&&
\left|\beta p_0+2\pi i\frac{a}{N}\right|_R
= \beta p_0+2\pi i\frac{a}{N} \\
&&
\left|\beta(p_0+k_0) +2\pi i\frac{a}{N}\right|_R
=
- \left(
\beta(p_0+k_0)+2\pi i\frac{a}{N}
\right).
\eea
Notice the difference from the previous case,
(\ref{plus1}) and (\ref{plus2}).
Then, (\ref{p=w}) results in a non-zero factor
\bea
&&
\frac{1}{2\w}
\sum_{a=1}^N
\frac{1}{e^{|\beta \w+2\pi i\frac{a}{N}|_R}-1}
\delta(k_0+2\w)
\frac{1}{e^{|-\beta \w+2\pi i\frac{a}{N}|_R}-1} \nn
&=&
\frac{1}{2\w}
\delta(k_0+2\w) 
\sum_{a=1}^N
\sum_{n_1=1}^\infty
e^{-n_1(\beta \w + 2\pi i\frac{a}{N})}
\sum_{n_2=1}^\infty
e^{-n_2(\beta \w - 2\pi i\frac{a}{N})}\nn
&=&
\frac{1}{2\w}
\delta(k_0+2\w) 
\sum_{n=1}^\infty
e^{-2n\beta \w}\nn
&=&
\frac{1}{2\w}
\delta(k_0+2\w) 
\frac{1}{e^{\beta (2\w)}-1}
\label{closedcut}.
\eea
The $a$-dependent
phase factors cancelled
when $n_1=n_2$ and
this gave the non-vanishing contribution.
The contribution
from the $p_0 = -\w$ case (\ref{p=-w})
can be calculated in the same way
and gives
$\frac{1}{2\w}\delta(k_0-2\w) 
\frac{1}{e^{\beta (2\w)}-1}$.
Together with (\ref{closedcut}),
this gives the factor
$\delta(k_0^2-(2\w)^2) \frac{1}{e^{\beta (2\w)}-1}$.
According to the AdS-CFT
dictionary,
the scaling dimension two of the operator 
$\tr \Phi_{(1)}^2$
corresponds
to a mass of the closed 
string state in units
of the inverse radius of the AdS,
which is $\w$ \cite{Witten:1998qj}.
Therefore,
this is the correct factor
for the temperature dependent
part of 
the type-1-type-1 propagator
of the {\em bulk}
closed string field theory
in the real time formalism.

\bibliography{crb}

\providecommand{\href}[2]{#2}\begingroup\raggedright\begin{thebibliography}{10}

\bibitem{Maldacena:2001kr}
J.~M. Maldacena, ``Eternal black holes in Anti-de-Sitter,'' { JHEP} {\bf 04}
  (2003) 021,
\href{http://www.arXiv.org/abs/hep-th/0106112}{{hep-th/0106112}}.

\bibitem{Maldacena:1997re}
J.~M. Maldacena, ``The large N limit of superconformal field theories and
  supergravity,'' { Adv. Theor. Math. Phys.} {\bf 2} (1998) 231--252,
\href{http://www.arXiv.org/abs/hep-th/9711200}{{hep-th/9711200}}.

\bibitem{Hawking:1982dh}
S.~W. Hawking and D.~N. Page, ``THERMODYNAMICS OF BLACK HOLES IN ANTI-DE SITTER
  SPACE,'' { Commun. Math. Phys.} {\bf 87} (1983)
577.

\bibitem{Witten:1998qj}
E.~Witten, ``Anti-de Sitter space and holography,'' { Adv. Theor. Math. Phys.}
  {\bf 2} (1998) 253--291,
\href{http://www.arXiv.org/abs/hep-th/9802150}{{hep-th/9802150}}.

\bibitem{Witten:1998zw}
E.~Witten, ``Anti-de Sitter space, thermal phase transition, and confinement in
  gauge theories,'' { Adv. Theor. Math. Phys.} {\bf 2} (1998) 505--532,
\href{http://www.arXiv.org/abs/hep-th/9803131}{{hep-th/9803131}}.

\bibitem{Matsubara:1955ws}
T.~Matsubara, ``A New approach to quantum statistical mechanics,'' { Prog.
  Theor. Phys.} {\bf 14} (1955)
351--378.

\bibitem{Gibbons:1976ue}
G.~W. Gibbons and S.~W. Hawking, ``ACTION INTEGRALS AND PARTITION FUNCTIONS IN
  QUANTUM GRAVITY,'' { Phys. Rev.} {\bf D15} (1977)
2752--2756.

\bibitem{Furuuchi:2005qm}
K.~Furuuchi, ``From free fields to AdS: Thermal case,'' { Phys. Rev.} {\bf D72}
  (2005) 066009,
\href{http://www.arXiv.org/abs/hep-th/0505148}{{hep-th/0505148}}.

\bibitem{Brigante:2005bq}
M.~Brigante, G.~Festuccia, and H.~Liu, ``Inheritance principle and
  Non-renormalization theorems at finite temperature,''
\href{http://www.arXiv.org/abs/hep-th/0509117}{{hep-th/0509117}}.

\bibitem{Furuuchi:2005eu}
K.~Furuuchi, ``Large N reductions and holography,''
\href{http://www.arXiv.org/abs/hep-th/0506183}{{hep-th/0506183}}.

\bibitem{Takahasi:1974zn}
Y.~Takahasi and H.~Umezawa, ``Thermo field dynamics,'' { Collect. Phenom.} {\bf
  2} (1975)
55--80.

\bibitem{Umezawa:1982nv}
H.~Umezawa, H.~Matsumoto, and M.~Tachiki, ``THERMO FIELD DYNAMICS AND CONDENSED
  STATES,''. Amsterdam, Netherlands: North-holland (1982) 591p.

\bibitem{Semenoff:1982ev}
G.~W. Semenoff and H.~Umezawa, ``FUNCTIONAL METHODS IN THERMO FIELD DYNAMICS: A
  REAL TIME PERTURBATION THEORY FOR QUANTUM STATISTICAL MECHANICS,'' { Nucl.
  Phys.} {\bf B220} (1983)
196--212.

\bibitem{Niemi:1983nf}
A.~J. Niemi and G.~W. Semenoff, ``FINITE TEMPERATURE QUANTUM FIELD THEORY IN
  MINKOWSKI SPACE,'' { Ann. Phys.} {\bf 152} (1984)
105.

\bibitem{Niemi:1983ea}
A.~J. Niemi and G.~W. Semenoff, ``THERMODYNAMIC CALCULATIONS IN RELATIVISTIC
  FINITE TEMPERATURE QUANTUM FIELD THEORIES,'' { Nucl. Phys.} {\bf B230} (1984)
181.

\bibitem{Landsman:1986uw}
N.~P. Landsman and C.~G. van Weert, ``REAL AND IMAGINARY TIME FIELD THEORY AT
  FINITE TEMPERATURE AND DENSITY,'' { Phys. Rept.} {\bf 145} (1987)
141.

\bibitem{Fidkowski:2003nf}
L.~Fidkowski, V.~Hubeny, M.~Kleban, and S.~Shenker, ``The black hole
  singularity in AdS/CFT,'' { JHEP} {\bf 02} (2004) 014,
\href{http://www.arXiv.org/abs/hep-th/0306170}{{hep-th/0306170}}.

\bibitem{Herzog:2002pc}
C.~P. Herzog and D.~T. Son, ``Schwinger-Keldysh propagators from AdS/CFT
  correspondence,'' { JHEP} {\bf 03} (2003) 046,
\href{http://www.arXiv.org/abs/hep-th/0212072}{{hep-th/0212072}}.

\bibitem{Balasubramanian:1998de}
V.~Balasubramanian, P.~Kraus, A.~E. Lawrence, and S.~P. Trivedi, ``Holographic
  probes of anti-de Sitter space-times,'' { Phys. Rev.} {\bf D59} (1999)
  104021,
\href{http://www.arXiv.org/abs/hep-th/9808017}{{hep-th/9808017}}.

\bibitem{Horowitz:1998xk}
G.~T. Horowitz and D.~Marolf, ``A new approach to string cosmology,'' { JHEP}
  {\bf 07} (1998) 014,
\href{http://www.arXiv.org/abs/hep-th/9805207}{{hep-th/9805207}}.

\bibitem{Israel:1976ur}
W.~Israel, ``Thermo field dynamics of black holes,'' { Phys. Lett.} {\bf A57}
  (1976)
107--110.

\bibitem{Kraus:2002iv}
P.~Kraus, H.~Ooguri, and S.~Shenker, ``Inside the horizon with AdS/CFT,'' {
  Phys. Rev.} {\bf D67} (2003) 124022,
\href{http://www.arXiv.org/abs/hep-th/0212277}{{hep-th/0212277}}.

\bibitem{Hartnoll:2005ju}
S.~A. Hartnoll and S.~Prem~Kumar, ``AdS black holes and thermal Yang-Mills
  correlators,''
\href{http://www.arXiv.org/abs/hep-th/0508092}{{hep-th/0508092}}.

\bibitem{Aharony:2003sx}
O.~Aharony, J.~Marsano, S.~Minwalla, K.~Papadodimas, and M.~Van~Raamsdonk,
  ``The Hagedorn / deconfinement phase transition in weakly coupled large N
  gauge theories,'' { Adv. Theor. Math. Phys.} {\bf 8} (2004) 603--696,
\href{http://www.arXiv.org/abs/hep-th/0310285}{{hep-th/0310285}}.

\bibitem{Aharony:2005bq}
O.~Aharony, J.~Marsano, S.~Minwalla, K.~Papadodimas, and M.~Van~Raamsdonk, ``A
  first order deconfinement transition in large N Yang- Mills theory on a small
  S**3,'' { Phys. Rev.} {\bf D71} (2005) 125018,
\href{http://www.arXiv.org/abs/hep-th/0502149}{{hep-th/0502149}}.

\bibitem{Sundborg:1999ue}
B.~Sundborg, ``The Hagedorn transition, deconfinement and N = 4 SYM theory,'' {
  Nucl. Phys.} {\bf B573} (2000) 349--363,
\href{http://www.arXiv.org/abs/hep-th/9908001}{{hep-th/9908001}}.

\bibitem{Polyakov:2001af}
A.~M. Polyakov, ``Gauge fields and space-time,'' { Int. J. Mod. Phys.} {\bf
  A17S1} (2002) 119--136,
\href{http://www.arXiv.org/abs/hep-th/0110196}{{hep-th/0110196}}.

\bibitem{'tHooft:1973jz}
G.~'t~Hooft, ``A PLANAR DIAGRAM THEORY FOR STRONG INTERACTIONS,'' { Nucl.
  Phys.} {\bf B72} (1974)
461.

\bibitem{Gopakumar:2003ns}
R.~Gopakumar, ``From free fields to AdS,'' { Phys. Rev.} {\bf D70} (2004)
  025009,
\href{http://www.arXiv.org/abs/hep-th/0308184}{{hep-th/0308184}}.

\bibitem{Gopakumar:2004qb}
R.~Gopakumar, ``From free fields to AdS. II,'' { Phys. Rev.} {\bf D70} (2004)
  025010,
\href{http://www.arXiv.org/abs/hep-th/0402063}{{hep-th/0402063}}.

\bibitem{Gopakumar:2005fx}
R.~Gopakumar, ``From free fields to AdS. III,'' { Phys. Rev.} {\bf D72} (2005)
  066008,
\href{http://www.arXiv.org/abs/hep-th/0504229}{{hep-th/0504229}}.

\bibitem{Gopakumar:2004ys}
R.~Gopakumar, ``Free field theory as a string theory?,'' { Comptes Rendus
  Physique} {\bf 5} (2004) 1111--1119,
\href{http://www.arXiv.org/abs/hep-th/0409233}{{hep-th/0409233}}.

\bibitem{Hata:1993gf}
H.~Hata and B.~Zwiebach, ``Developing the covariant Batalin-Vilkovisky approach
  to string theory,'' { Ann. Phys.} {\bf 229} (1994) 177--216,
\href{http://www.arXiv.org/abs/hep-th/9301097}{{hep-th/9301097}}.

\bibitem{Gubser:1998bc}
S.~S. Gubser, I.~R. Klebanov, and A.~M. Polyakov, ``Gauge theory correlators
  from non-critical string theory,'' { Phys. Lett.} {\bf B428} (1998) 105--114,
\href{http://www.arXiv.org/abs/hep-th/9802109}{{hep-th/9802109}}.

\bibitem{Leblanc:1987zj}
Y.~Leblanc, ``STRING FIELD THEORY AT FINITE TEMPERATURE,'' { Phys. Rev.} {\bf
  D36} (1987)
1780.

\bibitem{Berenstein:2002jq}
D.~Berenstein, J.~M. Maldacena, and H.~Nastase, ``Strings in flat space and pp
  waves from N = 4 super Yang Mills,'' { JHEP} {\bf 04} (2002) 013,
\href{http://www.arXiv.org/abs/hep-th/0202021}{{hep-th/0202021}}.

\bibitem{Abdalla:2005qs}
M.~C.~B. Abdalla, A.~L. Gadelha, and D.~L. Nedel, ``PP-wave light-cone free
  string field theory at finite temperature,'' { JHEP} {\bf 10} (2005) 063,
\href{http://www.arXiv.org/abs/hep-th/0508195}{{hep-th/0508195}}.

\bibitem{Mathur:1993tp}
S.~D. Mathur, ``Is the Polyakov path integral prescription too restrictive?,''
\href{http://www.arXiv.org/abs/hep-th/9306090}{{hep-th/9306090}}.

\bibitem{Abdalla:2004dg}
M.~C.~B. Abdalla, A.~L. Gadelha, and D.~L. Nedel, ``Closed string thermal torus
  from thermofield dynamics,'' { Phys. Lett.} {\bf B613} (2005) 213--220,
\href{http://www.arXiv.org/abs/hep-th/0410068}{{hep-th/0410068}}.

\bibitem{Sathiapalan:1986db}
B.~Sathiapalan, ``VORTICES ON THE STRING WORLD SHEET AND CONSTRAINTS ON TORAL
  COMPACTIFICATION,'' { Phys. Rev.} {\bf D35} (1987)
3277.

\bibitem{Kogan:1987jd}
Y.~I. Kogan, ``VORTICES ON THE WORLD SHEET AND STRING'S CRITICAL DYNAMICS,'' {
  JETP Lett.} {\bf 45} (1987)
709--712.

\bibitem{Atick:1988si}
J.~J. Atick and E.~Witten, ``THE HAGEDORN TRANSITION AND THE NUMBER OF DEGREES
  OF FREEDOM OF STRING THEORY,'' { Nucl. Phys.} {\bf B310} (1988)
291--334.

\bibitem{Kazakov:2000pm}
V.~Kazakov, I.~K. Kostov, and D.~Kutasov, ``A matrix model for the
  two-dimensional black hole,'' { Nucl. Phys.} {\bf B622} (2002) 141--188,
\href{http://www.arXiv.org/abs/hep-th/0101011}{{hep-th/0101011}}.

\bibitem{KalyanaRama:1998cb}
S.~Kalyana~Rama and B.~Sathiapalan, ``The Hagedorn transition, deconfinement
  and the AdS/CFT correspondence,'' { Mod. Phys. Lett.} {\bf A13} (1998)
  3137--3144,
\href{http://www.arXiv.org/abs/hep-th/9810069}{{hep-th/9810069}}.

\bibitem{Barbon:2001di}
J.~L.~F. Barbon and E.~Rabinovici, ``Closed-string tachyons and the Hagedorn
  transition in AdS space,'' { JHEP} {\bf 03} (2002) 057,
\href{http://www.arXiv.org/abs/hep-th/0112173}{{hep-th/0112173}}.

\bibitem{Barbon:2002nw}
J.~L.~F. Barbon and E.~Rabinovici, ``Remarks on black hole instabilities and
  closed string tachyons,'' { Found. Phys.} {\bf 33} (2003) 145--165,
\href{http://www.arXiv.org/abs/hep-th/0211212}{{hep-th/0211212}}.

\bibitem{McGreevy:2005ci}
J.~McGreevy and E.~Silverstein, ``The tachyon at the end of the universe,'' {
  JHEP} {\bf 08} (2005) 090,
\href{http://www.arXiv.org/abs/hep-th/0506130}{{hep-th/0506130}}.

\bibitem{Susskind:1993if}
L.~Susskind, L.~Thorlacius, and J.~Uglum, ``The Stretched horizon and black
  hole complementarity,'' { Phys. Rev.} {\bf D48} (1993) 3743--3761,
\href{http://www.arXiv.org/abs/hep-th/9306069}{{hep-th/9306069}}.

\bibitem{Hartle:1976tp}
J.~B. Hartle and S.~W. Hawking, ``PATH INTEGRAL DERIVATION OF BLACK HOLE
  RADIANCE,'' { Phys. Rev.} {\bf D13} (1976)
2188--2203.

\bibitem{Hartle:1983ai}
J.~B. Hartle and S.~W. Hawking, ``WAVE FUNCTION OF THE UNIVERSE,'' { Phys.
  Rev.} {\bf D28} (1983)
2960--2975.

\bibitem{Polyakov:1978vu}
A.~M. Polyakov, ``THERMAL PROPERTIES OF GAUGE FIELDS AND QUARK LIBERATION,'' {
  Phys. Lett.} {\bf B72} (1978)
477--480.

\bibitem{Hawking:1976ra}
S.~W. Hawking, ``BREAKDOWN OF PREDICTABILITY IN GRAVITATIONAL COLLAPSE,'' {
  Phys. Rev.} {\bf D14} (1976)
2460--2473.

\bibitem{Schwinger:1960qe}
J.~S. Schwinger, ``Brownian motion of a quantum oscillator,'' { J. Math. Phys.}
  {\bf 2} (1961)
407--432.

\bibitem{Keldysh:1964ud}
L.~V. Keldysh, ``Diagram technique for nonequilibrium processes,'' { Zh. Eksp.
  Teor. Fiz.} {\bf 47} (1964)
1515--1527.

\bibitem{Mathur:2003hj}
S.~D. Mathur, A.~Saxena, and Y.~K. Srivastava, ``Constructing 'hair' for the
  three charge hole,'' { Nucl. Phys.} {\bf B680} (2004) 415--449,
\href{http://www.arXiv.org/abs/hep-th/0311092}{{hep-th/0311092}}.

\bibitem{Festuccia:2005pi}
G.~Festuccia and H.~Liu, ``Excursions beyond the horizon: Black hole
  singularities in Yang-Mills theories. I,''
\href{http://www.arXiv.org/abs/hep-th/0506202}{{hep-th/0506202}}.

\bibitem{Hawking:1974sw}
S.~W. Hawking, ``Particle creation by black holes,'' { Commun. Math. Phys.}
  {\bf 43} (1975)
199--220.

\bibitem{Son:2002sd}
D.~T. Son and A.~O. Starinets, ``Minkowski-space correlators in AdS/CFT
  correspondence: Recipe and applications,'' { JHEP} {\bf 09} (2002) 042,
\href{http://www.arXiv.org/abs/hep-th/0205051}{{hep-th/0205051}}.

\end{thebibliography}\endgroup
\bibliographystyle{kazu} 

\end{document}